\documentclass[lettersize,journal]{IEEEtran}
\usepackage[fleqn]{amsmath}
\usepackage{booktabs}
\usepackage{longtable}
\usepackage[flushleft]{threeparttable}
\usepackage[flushleft]{threeparttablex}
\usepackage{multirow}
\usepackage{multicol}
\usepackage{amsmath,amssymb,amsfonts}
\usepackage{algorithm}
\usepackage{algpseudocode}
\usepackage{array}
\usepackage{textcomp}
\usepackage{stfloats}
\usepackage{url}
\usepackage{verbatim}
\usepackage{graphicx}
\usepackage{cite}
\usepackage{eurosym}
\usepackage{acronym}
\usepackage[mathscr]{eucal}
\usepackage{xcolor}
\usepackage{subfigure}
\usepackage{makecell} 

\usepackage{accents}
\newcommand{\ubar}[1]{\underaccent{\bar}{#1}}

\definecolor{dgreen}{rgb}{0.0, 0.5, 0.0}

\usepackage{nomencl}
\newcommand{\nomunit}[1]{%
\renewcommand{\nomentryend}{\hspace*{\fill}#1}}

\makenomenclature

\usepackage{etoolbox}
\renewcommand\nomgroup[1]{%
  \item[\bfseries
  \ifstrequal{#1}{G}{General Notation Concepts}{%
  \ifstrequal{#1}{I}{Indexes and Sets}{%
  \ifstrequal{#1}{P}{Parameters}{%
  \ifstrequal{#1}{V}{Variables}{%
  \ifstrequal{#1}{W}{Binary Variables}{}}}}}%
]}

\makeatletter
\g@addto@macro\normalsize{%
  \setlength{\abovedisplayskip}{2pt}
  \setlength{\belowdisplayskip}{2pt}
  \setlength{\abovedisplayshortskip}{2pt}
  \setlength{\belowdisplayshortskip}{2pt}
}

\makeatother

\graphicspath{{figs/}} 

\acrodef{dam}[DAM]{Day Ahead Market}
\acrodef{idm}[IDM]{Intra-Day Market}
\acrodef{srm}[SRM]{Secondary Reserve Market}
\acrodef{asm}[ASM]{Ancillary Service Market}
\acrodef{bam}[BAM]{Balancing Market}
\acrodef{sr}[SR]{Secondary Reserve}
\acrodef{pr}[PR]{Primary Reserve}
\acrodef{rto}[RTO]{Real-Time Operation}
\acrodef{rt}[RT]{Real-Time}
\acrodef{ess}[ESS]{Energy Storage System}
\acrodef{bess}[BESS]{Battery Energy Storage System}
\acrodef{res}[RES]{Renewable Energy Sources}
\acrodef{ndrs}[ND-RES]{Non-dispatchable Renewable Energy Sources}
\acrodef{drs}[D-RES]{Dispatchable Renewable Energy Sources}
\acrodef{csp}[CSP]{Concentrated Solar Power Plant}
\acrodef{vpp}[VPP]{Virtual Power Plant}
\acrodef{wf}[WF]{Wind Farm}
\acrodef{pv}[PV]{Photovoltaic}
\acrodef{opf}[OPF]{Optimal Power Flow}
\acrodef{pf}[PF]{Power Flow}
\acrodef{milp}[MILP]{Mixed Integer Linear Programming}
\acrodef{minlp}[MINLP]{Mixed Integer non-Linear Programming}
\acrodef{tso}[TSO]{Transmission System Operator}
\acrodef{so}[SO]{System Operator}
\acrodef{pcc}[PCC]{Point of Common Coupling}
\acrodef{ree}[REE]{Red El{\'e}ctrica de España}
\acrodef{afrr}[aFRR]{Automatic Frequency Restoration Reserve}
\acrodef{fcr}[FCR]{Frequency Containment Reserve}
\acrodef{rvpp}[RVPP]{Renewable-only Virtual Power Plant}
\acrodef{rVPP}[RVPP]{Renewable-only VPP}
\acrodef{gams}[GAMS]{General Algebraic Modeling System}
\acrodef{gdx}[.gdx]{GAMS Data eXchange}
\acrodef{picasso}[PICASSO]{Platform for the International Coordination of Automated Frequency Restoration and Stable System Operation}
\acrodef{so}[SO]{Stochastic Optimization}
\acrodef{ro}[RO]{Robust Optimization}
\acrodef{aro}[ARO]{Adaptive Robust Optimization}
\acrodef{saro}[SARO]{Stochastic Adaptive RO}
\acrodef{dro}[DRO]{Distributionally RO}

\acrodef{omie}[OMIE]{Spanish Market Operator}
\acrodef{agc}[AGC]{Automatic Generation Control}

\acrodef{mo}[MO]{Market Operator}

\acrodef{ccg}[C\&CG]{Column \& Constraint Generation}
\acrodef{rtm}[RTM]{Real-time Market}
\acrodef{ngm}[NGM]{Natural Gas Market}
\acrodef{pdf}[PDF]{Probability Density Function}
\acrodef{ev}[EV]{Electric Vehicle}
\acrodef{ptc}[PTC]{Parabolic Trough Collector}
\acrodef{pb}[PB]{Power Block}
\acrodef{sf}[SF]{Solar Field}
\acrodef{ts}[TS]{Thermal Storage}
\acrodef{hpa}[HPA]{Heat Purchase Agreement}
\acrodef{sos-2}[SOS-2]{Special Ordered Set of type 2}

\acrodef{eh}[EH]{Electric Heater}
\acrodef{ed}[ED]{Electric Demand}
\acrodef{td}[TD]{Thermal Demand}

\acrodef{es}[ESS]{Electrical Storage System}

\acrodef{tpp}[TPP]{Thermal Power Plant}
\acrodef{fd}[FD]{Flexible Demand}

\acrodef{igdt}[IGDT]{Information Gap Decision Theory}
\acrodef{cvar}[CVaR]{Conditional Value-at-Risk}
\acrodef{vab}[VaB]{Value-at-Best}
\acrodef{soc}[SoC]{State of Charge}

\acrodef{pvt}[PVT]{Photovoltaic-Thermal}

\acrodef{caes}[CAES]{Compressed Air Energy Storage}

\acrodef{ets}[ETS]{Emissions Trading System}
\acrodef{eua}[EUAs]{European Union Allowances}
\acrodef{cet}[CET]{Carbon Emission Trading}
\acrodef{ctm}[CTM]{Carbon Trading Market}
\acrodef{gos}[GOs]{Guarantees of Origin}
\acrodef{gct}[GCT]{Green Certificate Trading}
\acrodef{pso}[PSO]{Particle Swarm Optimization}
\acrodef{gt}[GT]{Gas Turbine}
\acrodef{p2g}[P2G]{Power-to-Gas}
\acrodef{cc}[CC]{Carbon Capture}
\acrodef{gs}[GS]{Gas Storage}
\acrodef{mt}[MT]{Micro Turbine}
\acrodef{cl}[CL]{Controllable Load}
\acrodef{woa}[WOA]{Whale Optimization Algorithm}
\acrodef{dr}[DR]{Demand Response}
\acrodef{chp}[CHP]{Combined Heat and Power}
\acrodef{cco}[CCO]{Chance-constrained Optimization}

\acrodef{rec}[REC]{Renewable Energy Certificate}
\acrodef{cer}[CER]{Carbon Emission Right}

\acrodef{ub}[UB]{Upper Bound}
\acrodef{lb}[LB]{Lower Bound}

\begin{document}

\title{Assessing Value of Renewable-based VPP Versus Electrical Storage: Multi-market Participation Under Different Scheduling Regimes and Uncertainties}

\author{Hadi Nemati,~\IEEEmembership{Student,~IEEE}, {Ignacio Egido}, Pedro S{\'a}nchez-Mart{\'i}n, {\'A}lvaro Ortega,~\IEEEmembership{Member,~IEEE \vspace{-8mm}}  
\thanks{The authors wish to thank Comunidad de Madrid for the financial support to PREDFLEX project (TEC-2024/ECO-287), through the R\&D activity programme Tecnologías 2024.}
}



\maketitle

\begin{abstract}
This paper compares the participation of \acp{rvpp} and grid‑scale \acp{es} in energy and reserve markets, evaluating their technical performance, market strategies, and economic outcomes. To ensure a fair comparison, scheduling is analyzed over representative sample days that capture seasonal operating regimes, and the associated uncertainties are explicitly modeled. Two‑stage robust optimization frameworks are employed: the \ac{rvpp} model addresses price, generation, and demand uncertainties, whereas the \ac{es} model considers price uncertainty only. In addition, an algorithm is proposed for sizing the \ac{es} so that its market performance matches that of the \ac{rvpp}. Simulations cover both \textit{favorable} and \textit{unfavorable} scenarios, reflecting seasonal energy limits for dispatchable resources, varying forecast errors for nondispatchable resources, and alternative uncertainty‑management strategies. The results provide operators with quantitative guidance on the relative value of each approach.


\end{abstract}

\vspace{-1.5em}

\section*{Nomenclature}
\vspace{-1.5em}


\setlength{\nomitemsep}{0.104cm}

\nomenclature[G, 01]{$\tilde{A}$}{Median of a forecast distribution \nomunit{}}
\nomenclature[G, 01]{$\hat{\tilde{A}}$, $\tilde{\check{A}}$}{Upper/lower bounds of forecast \nomunit{}}
\nomenclature[G, 01]{$\hat{A}$, $\check{A}$}{Positive/negative deviation from forecast \nomunit{}}
\nomenclature[G, 01]{$\bar{A}$, $\ubar{A}$}{Upper/lower bounds \nomunit{}}
\nomenclature[G, 01]{$A^{\uparrow}$, $A^{\downarrow}$}{Up/down regulation direction \nomunit{}}
\nomenclature[G, 01]{$A^{+}$, $A^{-}$}{Charging/discharging state for storage \nomunit{}}


\nomenclature[I, 01]{$c \in \mathscr{C}$}{Set of \acs{drs} \nomunit{}}
\nomenclature[I, 01]{$s \in \mathscr{S}$}{Set of \acsp{es} \nomunit{}}
\nomenclature[I, 01]{$d \in \mathscr{D}$}{Set of \acsp{fd} \nomunit{}}
\nomenclature[I, 01]{$r \in \mathscr{R}$}{Set of \acs{ndrs} \nomunit{}}
\nomenclature[I, 01]{$t \in \mathscr{T}$}{Set of time periods in each sample day \nomunit{}}
\nomenclature[I, 01]{$m \in \mathscr{M}$}{Set of daily load profiles \nomunit{}}
\nomenclature[I, 01]{$u \in \mathscr{U}$}{Set of units \nomunit{}}
\nomenclature[I, 02]{$\theta \in {\Theta}$}{Set of \acsp{csp} \nomunit{}}
\nomenclature[I, 03]{$\Xi^{\mathrm{R}/\mathrm{S}}$}{Set of decision variables of \acs{rvpp}/\acs{es} operator \nomunit{} \vspace{5pt}}

\nomenclature[P, 01]{$C_{u} \;$}{Operation costs of unit $u$ \nomunit{[€/MWh]}}
\nomenclature[P, 01]{$C_{u}^{SU/SD}$}{Start-up/shut-down costs of unit $u$ \nomunit{[€]}}

\nomenclature[P, 01]{$P_{u}$}{Electrical power capacity of unit $u$ \nomunit{[MW]}}

\nomenclature[P, 01]{$P_{r,t}$}{\acs{ndrs} $r$ production during period $t$ \nomunit{[MW]}}
\nomenclature[P, 01]{$P_{d,m,t}$}{\acs{fd} $d$ of profile $m$ consumption during period $t$ \nomunit{[MW]}}
\nomenclature[P, 01]{$P_{\theta,t}^{SF}$}{Thermal power of \acs{sf} of \acs{csp} $\theta$ during period $t$ \nomunit{[MW]}}



\nomenclature[P, 01]{$E_{u}$}{Electrical energy capacity of unit $u$ \nomunit{[MWh]}}
\nomenclature[P, 04]{$\eta_{s}$}{Electrical power efficiency of \acs{es} $s$ \nomunit{[\%]}}

\nomenclature[P, 01]{$K_{\theta}$}{Start up output multiplier of turbine of \acs{csp} $\theta$ \nomunit{[p.u.]}}
\nomenclature[P, 04]{$\eta_{\theta}$}{Thermal to electrical output efficiency of \acs{csp} $\theta$ \nomunit{[\%]}}


\nomenclature[P, 01]{$M \;$}{Big positive value \nomunit{[-]}}
\nomenclature[P, 04]{$ \Delta{t} \;$}{Duration of periods \nomunit{[hour]}}


\nomenclature[P, 03]{$\Gamma^{DA/SR}$}{Uncertainty budget of \acs{dam}/\acs{srm} price \nomunit{[-]} }
\nomenclature[P, 03]{$\Gamma_{u}$}{Uncertainty budget of unit $u$'s power output \nomunit{[-]}}

\nomenclature[P, 08]{$\lambda_t^{{DA}}$}{\acs{dam} price during period $t$ \nomunit{[€/MWh]}}
\nomenclature[P, 08]{$\lambda_t^{{SR,\uparrow (\downarrow)}}$}{ Up (down) \acs{srm} price during period $t$ \nomunit{[€/MW]}}

\nomenclature[V, 01]{$p_{u,t}$}{Electrical power of unit $u$ during period $t$ \nomunit{[MW]}}

\nomenclature[V, 01]{$p_t^{DA}$}{Electrical power traded by \acs{rvpp} during period $t$ \nomunit{[MW]}}

\nomenclature[V, 01]{$r_t^{SR}$}{Reserve traded by \acs{rvpp} during period $t$ \nomunit{[MW]}}

\nomenclature[V, 01]{$r_{u,t}$}{Reserve provided by unit $u$ during period $t$ \nomunit{[MW]}}

\nomenclature[V, 01]{$e_{s,t} \;$}{Electrical energy of \acs{es} $s$ during period $t$ \nomunit{[MWh]}}
\nomenclature[V, 02]{$\sigma_{s} \;$}{Share of \acs{es} $s$ capacity allocated for reserve \nomunit{[\%]}}

\nomenclature[V, 01]{$p_{\theta,t}^{{SF}/{TS}}$}{Thermal power of \acs{sf}/\acs{ts} of \acs{csp} $\theta$ during period $t$ \nomunit{[MW]}}

\nomenclature[V, 01]{$x_t^{DA}$}{Auxiliary variable of traded electrical energy \nomunit{[MWh]}}
\nomenclature[V, 01]{$x_{u,t}$}{Auxiliary variable of unit $u$'s power uncertainty \nomunit{[MW]}}


\nomenclature[V, 05]{$\xi_t^{DA/SR}$}{Dual variable of \acs{dam}/\acs{srm} price uncertainty \nomunit{[€]}}
\nomenclature[V, 05]{$\xi_{u,t}$}{Dual variable of unit $u$'s power uncertainty \nomunit{[MW]}}

\nomenclature[V, 03]{$\mu^{DA/SR}$}{Dual variable of \acs{dam}/\acs{srm} price uncertainty \nomunit{[€]}}
\nomenclature[V, 03]{$\mu_{u}$}{Dual variable of unit $u$'s power uncertainty \nomunit{[MW]}}


\nomenclature[V, 06]{$u_{u,t}$}{Binary variable of on/off status of unit $u$'s turbine \nomunit{[-]}}
\nomenclature[V, 06]{$v_{u,t}^{SU}$}{Binary variable of start-up status of unit $u$'s turbine \nomunit{[-]}}
\nomenclature[V, 06]{$u_{s,t}$}{Binary variable of charging/discharging state of \acs{es} $s$ \nomunit{[-]}}
\nomenclature[V, 06]{$u_{d,m}$}{Binary variable of selection of profile $m$ of \acs{fd} $d$ \nomunit{[-]}}

\nomenclature[V, 06]{$q_{u,t}$}{Binary variable of unit $u$'s power uncertainty \nomunit{[-]}}


\renewcommand{\nomname}{}
{\small
\printnomenclature[1.2cm]
}


\acresetall 
\section{Introduction}
\IEEEPARstart{T}{he} 
\ac{rvpp} functions as a unified operational entity that consolidates \ac{drs}, \ac{ndrs}, and \ac{fd} through digitalized infrastructure enabled by advanced forecasting, real-time monitoring, and decentralized energy management systems. By combining dispatchable assets such as hydro and biomass plants with variable resources like \ac{wf}, solar \ac{pv}, and \ac{csp}, alongside \ac{fd}, an \ac{rvpp} can offer enhanced operational flexibility and reliability. This enables participation in both \ac{dam} and \ac{srm}, where it can provide energy, up/down reserves, and demand-side response services through unified bidding strategies~\cite{yang2023optimal, shafiekhani2022optimal}. The ability to coordinate diverse units allows the \ac{rvpp} to smooth out renewable generation variability, optimize resource utilization, and manage power imbalances. Additionally, this aggregation framework enables inclusion of smaller or distributed energy resources that typically lack capacity to participate independently, thus promoting the integration of variable \ac{res} into wholesale electricity markets and contributing to grid reliability and economic performance~\cite{kaiss2025review}.

On the other hand, grid-scale \acp{es} are increasingly recognized as critical enablers of high \ac{res} penetration, offering fast response for both active and reactive power support to mitigate \ac{res} variability~\cite{kebede2022comprehensive}. Recent advances in lithium-ion technology have enhanced efficiency and scalability, while declining costs have accelerated deployment—evidenced by the Edwards \& Sanborn \ac{es} in California, which now operates at 3,287 MWh capacity~\cite{mortenson2024edwards}. Nonetheless, technical and environmental limitations remain. Lithium-ion \acp{es} generally maintain performance for approximately 2,000 to 5,000 charge-discharge cycles before experiencing significant degradation. Additionally, raw material extraction (e.g., lithium, cobalt) poses sustainability concerns, and recycling inefficiencies at end-of-life stages exacerbate environmental risks~\cite{nyamathulla2024review}. Overcoming these challenges is essential to ensure the long-term viability and environmental compatibility of \acp{es} in renewable-based power systems.

While both \ac{rvpp} and grid-scale \ac{es} possess the flexibility to participate in energy trading, arbitrage, and ancillary services such as reserve support~\cite{falabretti2023scheduling, kebede2022comprehensive}, their operational characteristics and energy availability differ significantly across scheduling regimes. For instance, \acp{es} are constrained by storage capacity and \ac{soc} for energy and reserve provision, whereas \ac{rvpp}s are limited by seasonal and weather-dependent resource availability across aggregated assets. These differences become especially pronounced under high uncertainty from market price volatility, renewable output fluctuations, and unpredictable load profiles~\cite{ghanuni2023risk}. As a result, the technical and economic performance of each solution varies depending on the market context and operational constraints. To accurately quantify and compare the value of \ac{rvpp} and \ac{es} participation, it is essential to develop advanced optimization frameworks that reflect these operational nuances. Such models must consider multiple scheduling regimes, technology-specific constraints, and stochastic representations of uncertainty. By integrating these aspects, a more realistic and comprehensive evaluation can be achieved, enabling stakeholders and decision makers to identify optimal strategies for resource coordination and market participation. 

The participation of \ac{rvpp}s in energy and reserve markets under multiple uncertainties to maximize profitability has been widely studied~\cite{yang2023optimal, shafiekhani2022optimal, nemati2025flexible, NEMATI2025136421, falabretti2023scheduling, zamani2016day, feng2025optimal, abbasi2019coordinated, rahimiyan2015strategic, xiong2024distributionally, naval2021virtual}. Table~\ref{table:Literature} categorizes existing works based on considered components, market structures, uncertainty modeling approaches, and scheduling horizons. The fundamental concept behind an \ac{rvpp} lies in coordinating controllable and dispatchable units to mitigate the stochastic nature of \ac{ndrs} generation, thereby ensuring operational viability and enabling the provision of multiple services across electricity markets. In this context, the integrating flexible resources such as hydro plant~\cite{yang2023optimal, naval2021virtual}, \ac{es}\cite{yang2023optimal, shafiekhani2022optimal, zamani2016day, feng2025optimal, rahimiyan2015strategic, xiong2024distributionally}, \ac{fd}\cite{shafiekhani2022optimal, nemati2025flexible, NEMATI2025136421, zamani2016day, feng2025optimal}, and \ac{ev}~\cite{falabretti2023scheduling, abbasi2019coordinated} is crucial for improving the controllability of intermittent \ac{res}. To address the complexities arising from multi-market participation and uncertainty, various optimization techniques have been employed, including \ac{milp}\cite{naval2021virtual}, \ac{so}\cite{yang2023optimal, falabretti2023scheduling, zamani2016day, abbasi2019coordinated}, \ac{ro}\cite{nemati2025flexible, NEMATI2025136421, rahimiyan2015strategic}, \ac{igdt}\cite{shafiekhani2022optimal}, data-driven~\cite{feng2025optimal}, and \ac{dro} approaches~\cite{xiong2024distributionally}. These methods are favored due to their ability to capture various technical constraints and efficiently handle multiple sources of uncertainty across scheduling horizons.

\begin{table*}[htbp]
  \centering
  \caption{Comparison of \ac{rvpp} approach in this paper and literature.}
\scriptsize
  \setlength{\tabcolsep}{3pt} 
  \renewcommand{\arraystretch}{.4} 
  \vspace{-1.3em}
  \begin{threeparttable}
  \begin{tabular}{@{}ccccccccccccccccccc@{}}  
    \toprule
    & \multicolumn{8}{c}{\textbf{Components}} 
    & \multicolumn{2}{c}{\textbf{Market}} 
    & \multicolumn{3}{c}{\textbf{Uncertainty}}  
    & \multicolumn{1}{c}{\textbf{Multiple scheduling}} 
    & \multicolumn{1}{c}{\textbf{\ac{rvpp} versus}}
    & \multicolumn{1}{c}{\textbf{Method}} \\

    \cmidrule(lr){2-9} \cmidrule(lr){10-11}  \cmidrule(lr){12-14}   
    \textbf{Ref.} & \textbf{\ac{pv}} & \textbf{\ac{wf}} & \textbf{Hydro} & \textbf{Biomass} & \textbf{\ac{csp}} & \textbf{\ac{es}} & \textbf{\ac{ev}} & \textbf{\ac{fd}} 
    & \textbf{Energy} & \textbf{Reserve}   
    & \textbf{Price} & \textbf{\ac{res}} & \textbf{Demand}  
    & \textbf{horizons} & \textbf{\ac{es} sizing} & \textbf{}   \\

    \cmidrule{1-17}

    \cite{yang2023optimal}
    & $\bullet$ & $\bullet$ & $\bullet$ &  &  & $\bullet$ &  &  & $\bullet$ & $\bullet$ 
    &  & $\bullet$ &  
    &  
    &
    &  \ac{so} \\ [0.2em]

        \cite{shafiekhani2022optimal}
    & $\bullet$ & $\bullet$ &  &  &  &  &  & $\bullet$
    & $\bullet$ &  
    & $\bullet$ & $\bullet$ &  
    &
    &
    & \ac{igdt} \\ [0.2em]


        \cite{nemati2025flexible, NEMATI2025136421}
    & $\bullet$ & $\bullet$ &  &  &  &  &  & $\bullet$
    & $\bullet$ & $\bullet$ 
    & $\bullet$ & $\bullet$  & $\bullet$ 
    &
    &
    & \ac{ro} \\ [0.2em]

        \cite{falabretti2023scheduling}
    & $\bullet$ &  &  &  &  & $\bullet$ & $\bullet$ & 
    & $\bullet$ &  
    &  & $\bullet$ & $\bullet$
    & $\bullet$
    &
    & \ac{so} \\ [0.2em]

        \cite{zamani2016day}
    & $\bullet$ & $\bullet$ &  &  &  & $\bullet$ &  & $\bullet$
    & $\bullet$ & $\bullet$ 
    & $\bullet$ & $\bullet$ & $\bullet$ 
    &
    &
    & Two-stage \ac{so} \\ [0.2em]

                \cite{feng2025optimal}
    & $\bullet$ & $\bullet$ &  &  &  & $\bullet$ &  & $\bullet$
    & $\bullet$ & $\bullet$ 
    &  & $\bullet$ &  
    &
    &
    & Data-driven \\ [0.2em]

                    \cite{abbasi2019coordinated}
    &  & $\bullet$ &  &  &  &  & $\bullet$ & 
    & $\bullet$ &  
    & $\bullet$ & $\bullet$ &  
    &
    &
    & \ac{so} \\ [0.2em]

                \cite{naval2021virtual}
    & $\bullet$ & $\bullet$ & $\bullet$ &  &  & $\bullet$ &  & 
    & $\bullet$ &  
    &  &  &  
    &
    &
    & \ac{milp} \\ [0.2em]

                    \cite{rahimiyan2015strategic}
    &  & $\bullet$ &  &  &  & $\bullet$ &  & $\bullet$
    & $\bullet$ &  
    & $\bullet$ & $\bullet$ &  
    &
    &
    & Two-stage \ac{ro} \\ [0.2em]

          \cite{xiong2024distributionally}
    &  & $\bullet$ &  &  & $\bullet$ &  &  & 
    & $\bullet$ &  
    &  & $\bullet$ &  
    &
    &
    & \ac{dro} \\ [0.2em]

    \textbf{This paper} & $\bullet$ & $\bullet$ & $\bullet$ & $\bullet$ & $\bullet$ & $\bullet$ &  & $\bullet$ 
    & $\bullet$ & $\bullet$
    & $\bullet$ & $\bullet$ & $\bullet$
    &  $\bullet$ 
    &  $\bullet$ 
    & Two-stage \ac{ro} \\

    \bottomrule
  \end{tabular}
  \end{threeparttable}
  \label{table:Literature}
 \vspace{-1.8em}
\end{table*}

While the above studies have contributed substantially to the field, most lack a holistic framework that considers diverse renewable technologies with varying scheduling characteristics and uncertainty profiles. Additionally, the literature offers limited insight into comparative profitability analyses of \ac{rvpp} and \ac{es} across energy and reserve markets. To address these gaps, this paper conducts a comprehensive technical and economic comparison of \ac{rvpp} and \ac{es}, considering multiple sources of uncertainty, seasonal variability in renewable output, and different scheduling regimes. A two-stage \ac{ro} framework is used to model the coordinated bidding and scheduling of an \ac{rvpp} comprising \ac{ndrs} and multiple dispatchable flexible resources. This model enables robust decision-making under uncertainty and supports multi-market participation across scheduling horizons. In addition, the analysis explores how individual technologies contribute to economic performance, providing deeper insight into the role of different configurations under practical market conditions. This study facilitates a clearer understanding of \ac{rvpp} and \ac{es} performance under different operational strategies, levels of uncertainty, and flexible technology configurations. The comparative assessment against different \ac{es}-based configurations offers valuable insights for system operators and decision-makers involved in deploying flexible, renewable-based assets.


Accordingly, this study makes the following contributions:

\begin{itemize}




\item {To evaluate the robustness and adaptability of \ac{rvpp} energy and reserve scheduling under generation and demand uncertainty, environmental variability in renewable output, and seasonal constraints of dispatchable units.}

\item {To compare the technical and economic performance of \ac{rvpp} and \ac{es} using an efficient solution algorithm, and to conduct a comparative analysis under different uncertainty-handling strategies and scheduling regimes.}

\item {To evaluate the economic value of unit technologies in electricity markets under various \ac{rvpp} and \ac{es} configurations.}

\end{itemize}

The remainder of this paper is structured as follows: Section~\ref{sec:Problem_description} outlines the problem scope. Section~\ref{sec:Formulation} presents the proposed formulation for comparing the \ac{rvpp} and \ac{es} in \ac{dam} and \ac{srm} participation. Section~\ref{sec:Case_Studies} discusses the case studies, and Section~\ref{sec:Conclusion} concludes the paper.

\section{Problem Description}
\label{sec:Problem_description}

\begin{figure}[!t]
    \centering
    \includegraphics[width=1\linewidth]{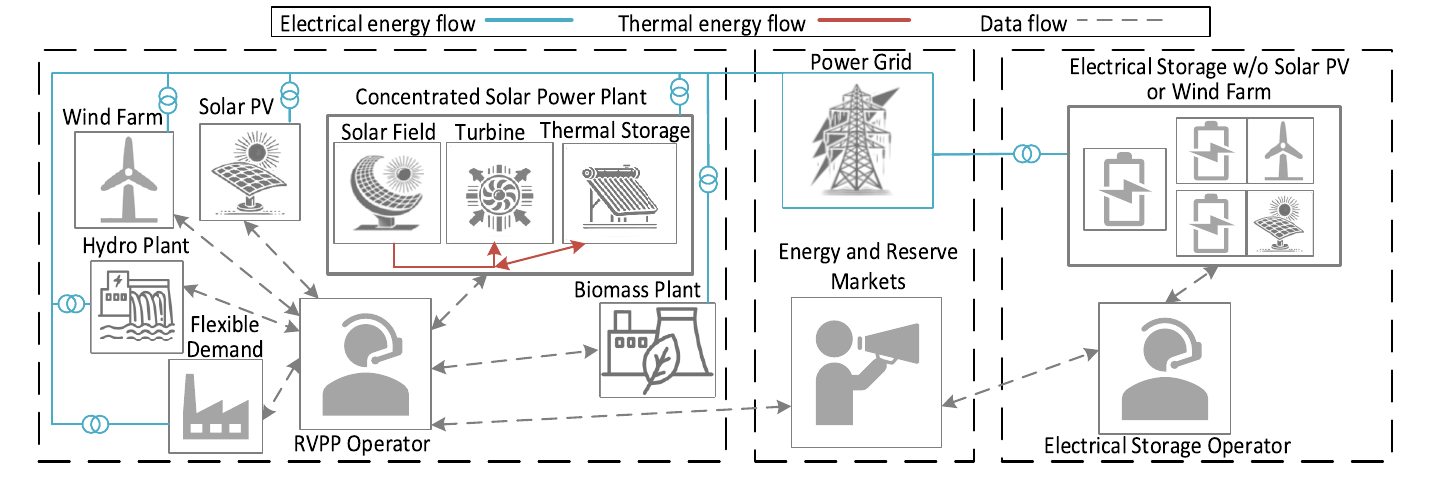}
    \vspace{-2.5em}
    \caption{The scheme of considered \ac{rvpp} and \ac{es}.}
     \label{fig:Scheme_RVPP}
     \vspace{-1.5em}
\end{figure}

Figure~\ref{fig:Scheme_RVPP} illustrates the schematic of the \ac{rvpp} and the \ac{es} participating in energy and reserve markets. By aggregating multiple \ac{res}, the \ac{rvpp} can coordinate internal dispatch and optimize market participation more effectively than individual units operating independently. The \ac{rvpp} operator uses forecasts of its \ac{ndrs} generation units and market prices, along with information on the availability of its dispatchable units and \ac{fd}, to determine an optimal bidding and scheduling strategy aimed at profit maximization. In this study, the \ac{rvpp} is modeled as a price-taker, submitting zero-price bids to reflect its relatively small scale compared to the overall electric grid. After receiving market-clearing results, the \ac{rvpp} operator communicates the dispatch decisions to its internal units. The \ac{es} problem is developed to determine the minimum required storage capacity needed to achieve economic performance equivalent to that of the \ac{rvpp}. The same market data is used for both problems to ensure a fair and comprehensive comparison between the two approaches. While standalone \ac{es} is modeled, it is also analyzed in joint configurations with \ac{ndrs} such as \ac{wf} and solar \ac{pv} to better reflect realistic market participation.

To further understand the impact of different technologies, multiple \ac{rvpp} configurations are evaluated by excluding key units (e.g., \ac{fd}, \ac{csp}, hydro) from \ac{rvpp}, enabling a detailed sensitivity analysis of their contributions. Accordingly, the corresponding \ac{es} setups for each configuration are analyzed. Given the seasonal variability of renewable energy availability and its effect on \ac{rvpp} profitability and scheduling, representative days for each season are modeled. Additionally, various uncertainty-handling strategies are considered to capture the impact of forecast variations in generation, consumption, and market prices. These factors enable a robust evaluation of both \ac{rvpp} and \ac{es} across diverse operational and market conditions. Two-stage optimization techniques are employed in the next section to address the market participation problems of \ac{rvpp} and \ac{es}, as they are well-suited for incorporating multiple uncertainties across different scheduling horizons.

\section{Formulation}
\label{sec:Formulation}
In this section, the deterministic models for \ac{rvpp} and \ac{es} participation in the electricity energy and reserve markets are first developed. Then, recognizing that price volatility, as well as generation and demand uncertainties, impact market outcomes, both models are extended to account for the corresponding uncertainties specific to each problem.

\subsection{Deterministic RVPP Problem}
\label{subsec:Deterministic_Formulation}
This section presents the deterministic \ac{rvpp} problem for joint participation in \ac{dam} and \ac{srm}. The objective is to maximize the \ac{rvpp}'s profit, subject to the technical constraints of the \ac{rvpp} units and the trading constraints.

\subsubsection{Objective Function}
\label{subsubsec:Objective_Function}
The \ac{rvpp} objective function~\eqref{RVPP: Obj_Deterministic} maximizes profit across the \ac{dam} and \ac{srm}, incorporating the operational expenses of its units. The terms represent, respectively: expected revenues from \ac{dam} bids; revenues from upward and downward \ac{srm} participation; operating costs of \ac{ndrs} and \acp{csp}; and operating, start-up, and shut-down costs of \ac{drs}.

\vspace{-1em}
\begingroup
\allowdisplaybreaks
\begin{align} \label{RVPP: Obj_Deterministic}
&\mathop {\max }\limits_{{\Xi ^{\mathrm{R}}}} \sum\limits_{t \in \mathscr{T}} { {\lambda _t^{DA}p_t^{DA}\Delta t   } } + \sum\limits_{t \in \mathscr{T}} {\left[ {{\lambda _t^{{SR, \uparrow}}r_t^{SR,\uparrow} }  +{\lambda _t^{{SR, \downarrow}}r_t^{SR,\downarrow} }  } \right]}
\nonumber \\& - \sum\limits_{t \in \mathscr{T}} {\sum\limits_{r \in \mathscr{R}} {C_rp_{r,t}\Delta t} } - \sum\limits_{t \in \mathscr{T}} {\sum\limits_{\theta \in \Theta} {C_{\theta}p_{\theta,t}\Delta t} } \nonumber \\& - \sum\limits_{t \in \mathscr{T}} {\sum\limits_{c \in \mathscr{C}} {\left[ {{C_c}p_{c,t}\Delta t +C_{c}^{SU} v_{c,t}^{SU} +C_{c}^{SD} v_{c,t}^{SD}} \right] }}  
\end{align}
\endgroup
\vspace{-.5em}

\subsubsection{Electrical Supply \& Demand Traded Constraints}
\label{subsubsec:Supply_Demand_Constraints}
The equality constraint for supply-demand balance among \ac{rvpp} units is formulated in~\eqref{cons: Supply-Demand1}. It accounts for all real-time reserve activation scenarios, including upward activation, downward activation, and no activation. To model these conditions, vectors $\boldsymbol{r}_{t}^{SR}=\{r_{t}^{SR,\uparrow}, -r_{t}^{SR, \downarrow},0\}$; $\boldsymbol{r}_{r,t}= \{r_{r,t}^{\uparrow},-r_{r,t}^{\downarrow},0 \}$; $\boldsymbol{r}_{\theta,t}= \{r_{\theta,t}^{\uparrow},-r_{\theta,t}^{\downarrow},0 \}$; $\boldsymbol{r}_{c,t}= \{r_{c,t}^{\uparrow},-r_{c,t}^{\downarrow},0 \}$; and $\boldsymbol{r}_{d,t}= \{r_{d,t}^{\uparrow},-r_{d,t}^{\downarrow},0\}$ are introduced for the \ac{rvpp}, \ac{ndrs}, \ac{csp}, \ac{drs}, and \ac{fd}, capturing the respective reserve states. Consequently,~\eqref{cons: Supply-Demand1} yields three distinct equations. 

\begingroup
\vspace{-.8em}
\allowdisplaybreaks
\begin{align}
    &\!\sum\limits_{r \in \mathscr{R}} \left[ p_{r,t} \!+ \boldsymbol{r}_{r,t} \right] \!+ \!\sum\limits_{\theta \in \Theta} \left[ p_{\theta,t} \!+ \boldsymbol{r}_{\theta,t} \right] \!+ \!\sum\limits_{c \in \mathscr{C}} \left[ p_{c,t} \!+ \boldsymbol{r}_{c,t} \right] 
    \nonumber \\& - \sum\limits_{d \in \mathscr{D}} \left[ p_{d,t} - \boldsymbol{r}_{d,t} \right] = p_{t}^{DA}+ \boldsymbol{r}_{t}^{SR}~;
    & 
    \forall t \label{cons: Supply-Demand1}
%
%
%
 %
\end{align}
\vspace{-.5em}
\endgroup

\subsubsection{Dispatchable Unit Constraints}
\label{subsubsec:TPP_Constraints}

Constraints~\eqref{Deterministic: DRES1}-\eqref{Deterministic: DRES2} bound \ac{drs} production considering reserve provision. 
Minimum up/down time constraints, following the formulation in~\cite{carrion2006computationally}, are omitted here for brevity. 
Daily energy limits due to seasonal regulations are addressed in~\eqref{Deterministic: DRES13}.

\begingroup
\allowdisplaybreaks
\begin{subequations}
\vspace{-.9em}
\begin{align}
    &p_{c,t} + r_{c,t}^{\uparrow} \leq \bar P_{c} u_{c,t}~; 
    &
    \forall c, t \label{Deterministic: DRES1} \\
    &\ubar P_{c} u_{c,t} \leq p_{c,t} - r_{c,t}^{\downarrow}~; 
    &
    \forall c, t \label{Deterministic: DRES2} \\
    %
    %
    %
    %
    %
    %
    %
    %
       %
        %
    %
    &\sum\limits_{t \in \mathscr{T}} \left[{p_{c,t}\Delta t} + r_{c,t}^{\uparrow} \right] \le \bar E_{c}~;
    &
    \forall c \label{Deterministic: DRES13} 
    %
    %
    \end{align}    
\label{Deterministic: DRES}
\end{subequations}
    \vspace{-1em}
\endgroup

\subsubsection{Non-dispatchable Unit Constraints}
\label{subsubsec:RES_Constraints}

The constraints for \ac{ndrs} are defined in~\eqref{RVPP: NDRES}. Equations~\eqref{cons: NDRES1} and~\eqref{cons: NDRES2} specify the bounds on energy and reserve outputs based on the fixed value of the associated uncertainty~\cite{zhang2022frequency}. 

\begingroup
\allowdisplaybreaks
\begin{subequations}
\vspace{-.8em}
\begin{align}
    & p_{r,t}+r_{r,t}^{\uparrow} \leq P_{r,t}~; 
    & 
    \forall r,t \label{cons: NDRES1} \\ 
    & \ubar P_{r} \le p_{r,t}-r_{r,t}^{\downarrow}~; 
    & 
    \forall r,t  \label{cons: NDRES2} 
%
%
    \end{align}
\label{RVPP: NDRES}
\end{subequations}
\vspace{-1em}
\endgroup

\subsubsection{Concentrated Solar Power Plant Constraints}
\label{subsubsec:STU}


The transformation of thermal to electrical energy in the CSP turbine is represented by~\eqref{Deterministic: STU}~\cite{ortega2022modeling}. Constraint~\eqref{Deterministic: STU1} defines the allowable range of thermal output from the \ac{sf}, given a fixed solar irradiation parameter. Equation~\eqref{Deterministic: STU2} integrates the thermal power dispatched to the turbine from the \ac{sf}, the charging and discharging of \ac{ts}, and turbine startup losses. The conversion efficiency of the turbine is represented by the efficiency parameter ${\eta_{\theta}}$. Constraints~\eqref{Deterministic: STU3}-\eqref{Deterministic: STU4} bound the \ac{csp}'s electrical output and reserve according to maximum/minimum limits and the turbine's binary commitment status $u_{\theta,t}$. Commitment status and minimum up/down time constraints are based on~\cite{carrion2006computationally}. The \ac{ts} constraints follow the formulation for \acp{es} described in Section~\ref{subsubsec:ES_Operation} and are omitted here for brevity. 

\begingroup
\allowdisplaybreaks
\begin{subequations}
\vspace{-.5em}
\begin{align}
    & 0 \leq p_{\theta,t}^{SF} \leq {P}_{\theta,t}^{SF}~; 
    & \forall \theta, t \label{Deterministic: STU1} \\
    & \frac{p_{\theta,t}}{\eta_{\theta}} 
    = p_{\theta,t}^{SF} 
    + p_{\theta,t}^{TS, -} - p_{\theta,t}^{TS, +} 
    - K_{\theta} v_{\theta,t}^{SU} \bar P_{\theta}~; 
    &\forall \theta, t \label{Deterministic: STU2} \\
    &p_{\theta,t} + r_{\theta,t}^{\uparrow} \leq \bar P_{\theta} u_{\theta,t}~; 
    &\forall \theta, t \label{Deterministic: STU3} \\
    &\ubar P_{\theta} u_{\theta,t} \leq p_{\theta,t} - r_{\theta,t}^{\downarrow}~; 
    &\forall \theta, t \label{Deterministic: STU4}
    %
    %
    %
    \end{align} 
\label{Deterministic: STU}
\end{subequations}
\vspace{-1em}
\endgroup

\subsubsection{Flexible Demand Constraints}
\label{subsubsec:Demand_Constraints}


The deterministic constraints governing the \acp{fd} are formulated in~\eqref{RVPP: Demand}~\cite{ortega2022modeling}. Constraint~\eqref{cons: Demand1} defines the minimum consumption level for \acp{fd} during each time period, considering the possibility of selecting different demand profiles. Constraint~\eqref{cons: Demand2} enforces the selection of exactly one demand profile among multiple candidates. The permissible operating range for FDs, encompassing both energy consumption and reserve provision, is bounded by~\eqref{cons: Demand3} and~\eqref{cons: Demand4}. 

\begingroup
\vspace{-.5em}
\allowdisplaybreaks
\begin{subequations}
\begin{align}
    & p_{d,t} \geq \sum_{m \in  \mathscr{M}} \left[ {P}_{d,m,t} u_{d,m}  \right]~;
    & 
    \forall d,t \label{cons: Demand1}\\
    & \sum_{m \in \mathscr{M}} u_{d,m} = 1~; 
    & \forall d \label{cons: Demand2} \\
    & \ubar P_{d} \le p_{d,t} - r_{d,t}^{\uparrow}~;
    &
    \forall d,t  \label{cons: Demand3} \\
    & p_{d,t} + r_{d,t}^{\downarrow} \le \bar P_{d}~;
    & 
    \forall d,t  \label{cons: Demand4}
%
%
%
%
%
%
\end{align}
\label{RVPP: Demand}
\end{subequations}
\vspace{-.5em}
\endgroup

\vspace{-1.3em}
\subsection{Electrical Storage System Problem}
\label{subsec:ES_Problem}
This section focuses on identifying the \ac{es} capacity required to match the operational and market performance of an \ac{rvpp} across energy and reserve markets. First, the operational modeling of \ac{es} participation in the \ac{dam} and \ac{srm} is detailed. Then, an algorithm for sizing the \ac{es} to achieve comparable economic profitability to the \ac{rvpp} is introduced.


\subsubsection{Electrical Storage Operation Constraints}
\label{subsubsec:ES_Operation}

The \ac{es} formulation in~\eqref{Deterministic: ES} integrates energy trading and reserve provision. The objective~\eqref{Deterministic: ES1} maximizes \ac{dam} and \ac{srm} profits minus operation and degradation costs. Constraints~\eqref{Deterministic: ES2}–\eqref{Deterministic: ES5} manage \ac{es} charging $(+)$ and discharging $(-)$, considering upward and downward reserves in both states, with the binary variable $u_{s,t}$ indicating the \ac{es} state. Output power and reserve are determined in~\eqref{Deterministic: ES6}–\eqref{Deterministic: ES8}, based on contributions from the charging and discharging states. The \ac{soc} of the \ac{es} is modeled in~\eqref{Deterministic: ES9}, while~\eqref{Deterministic: ES10} ensures daily energy balance by maintaining consistent initial and final \ac{soc}. Variables $\sigma_{s}^{\uparrow}$ and $\sigma_{s}^{\downarrow}$ in~\eqref{Deterministic: ES11}–\eqref{Deterministic: ES13} represent the shares of \ac{es} capacity allocated for upward and downward regulation, respectively~\cite{ortega2022modeling}. 

\begingroup
\allowdisplaybreaks
\begin{subequations}
\vspace{-.5em}
\begin{align}
    &\mathop {\max}\limits_{{\Xi^{\mathrm{S}}}} 
    \sum\limits_{t \in \mathscr{T}} \sum\limits_{s \in \mathscr{S}} {\left[ {\lambda _t^{DA}p_{s,t}\Delta t   } + {{\lambda _t^{{SR, \uparrow}}r_{s,t}^{\uparrow} } +{\lambda _t^{{SR, \downarrow}}r_{s,t}^{\downarrow} }  } - C_sp_{s,t}^{-} \right]} 
    \label{Deterministic: ES1}
    \end{align}
    \vspace{-.5cm}
\begin{align}
    \nonumber\text{st.} \\
    &\ubar {P}_{s}^{+} u_{s,t} \leq p_{s,t}^{+} - r_{s,t}^{+,\uparrow}~; 
    &\forall s, t \label{Deterministic: ES2} \\
    & p_{s,t}^{+} + r_{s,t}^{+,\downarrow} \leq \bar {P}_{s}^{+} u_{s,t}~;
    &\forall s, t \label{Deterministic: ES3} \\
    &p_{s,t}^{-} + r_{s,t}^{-,\uparrow} \leq \bar {P}_{s}^{-} \left( 1 - u_{s,t} \right)~; 
    &\forall s, t \label{Deterministic: ES4} \\
    &\ubar {P}_{s}^{-} \left( 1 - u_{s,t} \right) \leq p_{s,t}^{-} - r_{s,t}^{-,\downarrow}~; 
    &\forall s, t \label{Deterministic: ES5} \\
    &p_{s,t} = p_{s,t}^{-} - p_{s,t}^{+}~; 
    &\forall s, t \label{Deterministic: ES6} \\
    &r_{s,t}^{\uparrow} = r_{s,t}^{+,\uparrow} + r_{s,t}^{-,\uparrow}~; 
    &\forall s, t \label{Deterministic: ES7} \\
    &r_{s,t}^{\downarrow} = r_{s,t}^{+,\downarrow} + r_{s,t}^{-,\downarrow}~; 
    &\forall s, t \label{Deterministic: ES8} \\
    & e_{s,t} = e_{s,t-1} + p_{s,t}^{+} \eta_s^{+} \Delta t - \frac{p_{s,t}^{-} \Delta t}{\eta_{s}^{-}}~;
    &\forall s, t \backslash\{1\} \label{Deterministic: ES9} \\
    & e_{s,1} = e_{s,t=T}~;
    &\forall s \label{Deterministic: ES10} \\
    & \sum_{t \in \mathscr{T}} \frac{r_{s,t}^{\uparrow} \Delta t} {\eta_s^{-}} \leq \sigma_{s}^{\uparrow} \left( \bar E_s - \ubar E_s \right)~; 
    &\forall s \label{Deterministic: ES11} \\
    & \sum_{t \in \mathscr{T}} r_{s,t}^{\downarrow} \eta_s^{+} \Delta t \leq \sigma_{s}^{\downarrow} \left( \bar E_s - \ubar E_s \right)~; 
    &\forall s \label{Deterministic: ES12}
    \end{align}
    \vspace{-.2cm}
    \begin{align}
    &\ubar E_s \!\!+ \sigma_s^{\downarrow} \left( \bar E_s \!\!- \ubar E_s \right) \!\!\leq e_{s,t} \!\!\leq \bar E_{s} \!\!- \sigma_{s}^{\downarrow} \left( \bar E_{s} \!\!- \ubar E_{s} \right);& 
    \forall s, t \label{Deterministic: ES13}   
    %
\end{align}
\label{Deterministic: ES}
\end{subequations}
\vspace{-.6em}
\endgroup

\vspace{-.5em}
\subsubsection{Electrical Storage Sizing}
\label{subsubsec:ES_Sizing}

To determine the \ac{es} capacity required to match the economic performance as the \ac{rvpp}, both the \ac{rvpp} optimization model~\eqref{RVPP: Obj_Deterministic}–\eqref{RVPP: Demand} and the \ac{es} operation model~\eqref{Deterministic: ES} are employed. First, the profitability of the \ac{rvpp} in the \ac{dam} and \ac{srm} is compared with the aggregated operation profit of the individual units under the assumption that each participates independently in the markets. This difference is then used as the \ac{lb} in the \ac{es} optimization model~\eqref{Deterministic: ES}. The parameters of the \ac{es} are iteratively updated by incrementally adding modules of the \ac{es} until the operation profit of the \ac{es} reaches or exceeds the profitability level of the \ac{rvpp}. The steps for determining appropriate \ac{es} capacity are illustrated in Algorithm~\ref{alg:ES_capacity}.

\begin{algorithm}[t!]

\caption{ \small{Size of \ac{es} to match \ac{rvpp} economic performance.}}
\label{alg:ES_capacity}
 \footnotesize
\begin{algorithmic}[1]
\State \textbf{Input:} Technical parameters and forecast data for \ac{rvpp} and \ac{es}.
\State Solve the \ac{rvpp} optimization problem~\eqref{RVPP: Obj_Deterministic}-\eqref{RVPP: Demand}.
\State Compare \ac{rvpp} profitability with individual unit participation and set this value as the \ac{lb} of the \ac{es} problem~\eqref{Deterministic: ES}.
\Repeat
    \State Solve the \ac{es} optimization problem~\eqref{Deterministic: ES}.
    \If{problem~\eqref{Deterministic: ES} is not feasible}
        \State Update \ac{es} input parameters $\ubar {P}_{s}^{+}$, $\bar {P}_{s}^{+}$, $\ubar {P}_{s}^{-}$, $\bar {P}_{s}^{-}$, $\ubar E_s$, $\bar E_s$ by adding one module.
    \EndIf
\Until{problem~\eqref{Deterministic: ES} becomes feasible}

\State \textbf{Output:} Final \ac{es} capacity and its optimal scheduling and operation.
\end{algorithmic}
\end{algorithm}



\vspace{-.5em}
\subsection{Robust RVPP Problem}
\label{subsec:RVPP_Robust}

The robust \ac{rvpp} problem considers uncertainties in \ac{dam} and \ac{srm} prices, \ac{ndrs} and \acp{csp} generation, and \acp{fd} consumption. A flexible two-stage \ac{ro} model is developed and reformulated as a single-level \ac{milp} to handle uncertainties.

\subsubsection{Two-stage Problem}
\label{subsubsec:RVPP_Two-Stage}

In the first stage of model~\eqref{Two-Stage_Robust}, the \ac{rvpp} operator maximizes its objective function~\eqref{Robust1: Obj_Robust}, which mirrors the deterministic objective function~\eqref{RVPP: Obj_Deterministic}. $O^{\mathrm{R}}$ refers to the terms in the objective function of the deterministic \ac{rvpp} problem that are not affected by uncertainty. In the second stage, uncertainties negatively impact the electricity prices in the \ac{dam} and \ac{srm}, as represented by the minimization part in the objective function. Uncertainties are also modeled to potentially reduce the electrical output of \ac{ndrs} and the thermal output of \acp{csp}, while increasing the consumption of \acp{fd}, as expressed in constraints~\eqref{Robust1: NDRES1_Robust}-\eqref{Robust1: Demand1_Robust}. Notably, unlike the deterministic formulation, the uncertainty sets $\{ \lambda_t^{DA}, \lambda_t^{{SR,\uparrow}}, \lambda_t^{{SR, \downarrow}} \}$ and $\{ {P}_{r,t}, P_{\theta,t}^{SF}, {P}_{d,t}\}$ (index $m$ in ${P}_{d,m,t}$ from~\eqref{cons: Demand1} is omitted for simplicity) now include second-stage decision variables, which were treated as fixed parameters in the deterministic model. Constraints from Section~\ref{subsec:Deterministic_Formulation} unaffected by uncertainty are defined by $C^{\mathrm{R}}$ in~\eqref{Robust1: Other_Cons_Det}\footnote{Deterministic constraints include: \eqref{cons: Supply-Demand1}, \eqref{Deterministic: DRES}, \eqref{cons: NDRES2}
, \eqref{Deterministic: STU2}-\eqref{Deterministic: STU4}, \eqref{cons: Demand2}-\eqref{cons: Demand4}.}.

\begingroup
\vspace{-.8em}
\allowdisplaybreaks
\begin{subequations}
\begin{align} \label{Robust1: Obj_Robust}
&\mathop {\max}\limits_{{\Xi ^{\mathrm{R}}}} \scalebox{1}{\Bigg\{ } \mathop {\min}\limits_{\{ \lambda _t^{DA}, \lambda _t^{{SR, \uparrow}}, \lambda _t^{{SR, \downarrow}} \}} \scalebox{.8}{\Bigg\{ } \sum\limits_{t \in \mathscr{T}} { {\lambda _t^{DA}p_t^{DA}\Delta t   } } \nonumber \\&+ \sum\limits_{t \in \mathscr{T}} {\left[ {{\lambda _t^{{SR, \uparrow}}r_t^{SR,\uparrow} }  + {\lambda _t^{{SR, \downarrow}}r_t^{SR,\downarrow} }  } \right]} -O^{\mathrm{R}} \scalebox{.8}{\Bigg\} } \scalebox{1}{\Bigg\} } 
\end{align}
\begin{align}
    \nonumber\text{st.} \\
    & p_{r,t}+r_{r,t}^{\uparrow} \leq \mathop {\min}\limits_{{{P}_{r,t}}} \left\{ P_{r,t} \right\}~;  
     & \forall r,t \label{Robust1: NDRES1_Robust} \\ 
    & p_{\theta,t}^{SF} \leq \mathop {\min}\limits_{{{P}_{\theta,t}^{SF}}} \left\{ {P}_{\theta,t}^{SF} \right\}~;
     & \forall \theta, t \label{Robust1: STU1_Robust} \\
    & p_{d,t} \geq - \mathop {\min}\limits_{{{P}_{d,t}}} \left\{ - {P}_{d,t}  \right\}~; 
     & \forall d,t   \label{Robust1: Demand1_Robust}\\
    &C^{\mathrm{R}} \leq 0;
    &   
    \label{Robust1: Other_Cons_Det}
\end{align}
\label{Two-Stage_Robust}
\end{subequations}
\endgroup
\vspace{-1em}

\subsubsection{Single-level Reformulation}
\label{subsubsec:MILP_Formulation}

The single-level \ac{milp} formulation~\eqref{Single-level_MILP} is derived by applying the strong duality principle to the original \ac{ro} problem~\eqref{Two-Stage_Robust}~\cite{nemati2025flexible}. The uncertainty bounds in the optimization problem~\eqref{Single-level_MILP} are governed by \textit{uncertainty budget} parameters. Each budget is an integer from 0 to 24 for each hour of the sample day, allowing the conservatism level to vary from optimistic to pessimistic. The objective function~\eqref{MILP: Obj} captures the worst-case impact of uncertainties in \ac{dam} and \ac{srm} prices. 
Asymmetric price deviations in the \ac{dam} are modeled via constraint~\eqref{MILP_asymetric1}. Dual representations of the \ac{dam} and \ac{srm} price uncertainties are formulated in constraints~\eqref{MILP: DAMprice}–\eqref{MILP: downSRMprice}. Uncertainty associated with the electrical production of \acp{ndrs} is addressed through constraints~\eqref{MILP: NDRES1}–\eqref{MILP: NDRES4}. 
Similarly, uncertainties in \acp{csp} thermal production and \acp{fd} consumption are handled by analogous constraints, which are omitted for brevity. 
The remaining deterministic constraints are included in~\eqref{MILP: Other_Cons_Det}.

\begingroup
\vspace{-1em}
\begin{subequations}
\allowdisplaybreaks
\begin{align} \label{MILP: Obj}
&\mathop {\max} \limits_{{\Xi^{\mathrm{R}}}} \!\!\scalebox{1}{\Bigg\{} 
\sum\limits_{t \in \mathscr{T}} {\left[{\tilde{\lambda}_t^{DA}p_t^{DA}\Delta t   }\!+ {{ {\hat{\tilde{\lambda}}_t^{{SR, \uparrow}}r_t^{SR,\uparrow} }  \!+{\hat{\tilde{\lambda}}_t^{{SR, \downarrow}}r_t^{SR,\downarrow} }  } }\right]} \!- O^{\mathrm{R}} 
\nonumber \\&
 - \Gamma^{DA} \mu^{DA} - \sum\limits_{t \in \mathscr{T}} {\xi}^{DA}_t - \Gamma^{SR,\uparrow} \mu^{SR,\uparrow} - \Gamma^{SR,\downarrow} \mu^{SR,\downarrow}   \nonumber \\&- \sum\limits_{t \in \mathscr{T}} \left[ {\xi}^{SR,\uparrow}_t+ {\xi}^{SR, \downarrow}_t\right]  \scalebox{1}{\Bigg\}}
\end{align}
\vspace{-.5cm}
\begin{align}
    \nonumber\text{st.} \\
    &-\frac{\check{\lambda}_t^{DA}}{\hat{\lambda}_t^{DA}} {x_t}^{DA} \le {p_t}^{DA} \Delta t \le {x_t}^{DA}~; 
     &   
    \forall t \label{MILP_asymetric1} \\
    &{\mu^{DA}} + \xi_t^{DA} \ge  \check{\lambda}_t^{DA}x_t^{DA}~; 
     &    
    \forall t \label{MILP: DAMprice} \\
    &{\mu^{SR,\uparrow}} + \xi_t^{SR,\uparrow} \ge ~\check{\lambda}_t^{SR,\uparrow} r_t^{SR,\uparrow}~; 
     &   
    \forall t \label{MILP: upSRMprice} \\
    &{\mu^{SR,\downarrow}} + \xi_t^{SR,\downarrow} \ge ~\check{\lambda}_t^{SR,\downarrow} r_t^{SR,\downarrow}~; 
     &  
    \forall t \label{MILP: downSRMprice} \\
    & p_{r,t}+r_{r,t}^{\uparrow} \leq \hat{\Tilde{P}}_{r,t} 
    - x_{r,t}~;  
     & \forall r,t \label{MILP: NDRES1} \\ 
    &\mu_{r} \!+ {\xi}_{r,t} \!- M (1 \!- q_{r,t}) \!\leq x_{r,t} \!\leq \!M q_{r,t};  
    & \forall r,t \label{MILP: NDRES2} \\ 
    &\mu_{r} + \xi_{r,t} \ge  \check{P}_{r,t}~; 
     &   
    \forall r, t \label{MILP: NDRES3} \\
    & \sum_{t} q_{r,t} = \Gamma_{r}~; 
     &
    \forall r \label{MILP: NDRES4} \\
    %
    %
    %
     %
    %
    %
    %
     %
        %
 %
    %
    %
    & C^{\mathrm{R}} \leq 0;
    &    
    \label{MILP: Other_Cons_Det}
\end{align}
\label{Single-level_MILP}
\end{subequations}
\vspace{-2em}
\endgroup

\subsection{Robust ESS Problem}
\label{subsec:ESS_Robust}

The robust \ac{es} model addresses price uncertainty in the \ac{dam} and \ac{srm}. Adopting the methodology outlined in Section~\ref{subsec:RVPP_Robust}, the \ac{es} problem is recast as the single‑level \ac{milp} given in~\eqref{Single-level_MILP_ES}. The symbols $O^{\mathrm{S}}$ in~\eqref{MILP_ES: Obj} denotes the objective terms in the deterministic \ac{es} model~\eqref{Deterministic: ES} that remain unaffected by uncertainty. In the objective function~\eqref{MILP_ES: Obj}, the first line captures all deterministic components, whereas the second line introduces dual variables that penalize the worst‑case realizations of \ac{dam} and \ac{srm} price uncertainty. The corresponding dual feasibility conditions, which distinguish between the charging and discharging modes of the \ac{es}, are specified in~\eqref{MILP_ES: DAMprice}–\eqref{MILP_ES: downSRMprice}. The remaining operational constraints unaffected by uncertainty are consolidated in~\eqref{MILP_ES: Other_Cons_Det}.

\vspace{-1em}
\begingroup
\begin{subequations}
\allowdisplaybreaks
\begin{align} \label{MILP_ES: Obj}
&\mathop {\max}\limits_{{\Xi^{\mathrm{S}}}} \scalebox{1}{\Bigg\{} 
\sum\limits_{t \in \mathscr{T}} \sum\limits_{s \in \mathscr{S}} {\left[ {\tilde{\lambda}_t^{DA}p_{s,t}\Delta t} + {{ {\hat{\tilde{\lambda}}_t^{{SR, \uparrow}}r_{s,t}^{\uparrow} }  +{\hat{\tilde{\lambda}}_t^{{SR, \downarrow}}r_{s,t}^{\downarrow} }  } }\right]} - O^{\mathrm{S}} 
\nonumber \\&
 - \Gamma^{DA} \mu^{DA} - \sum\limits_{t \in \mathscr{T}} {\xi}^{DA}_t - \Gamma^{SR,\uparrow} \mu^{SR,\uparrow} - \Gamma^{SR,\downarrow} \mu^{SR,\downarrow} \nonumber \\&  - \sum\limits_{t \in \mathscr{T}} \left[ {\xi}^{SR,\uparrow}_t+ {\xi}^{SR, \downarrow}_t\right]  \scalebox{1}{\Bigg\}}
\end{align}
\begin{align}
    \nonumber\text{st.} \\
    &\!\!{\mu^{DA}} \!\!+ \xi_t^{DA} \!\!\ge \check{\lambda}_t^{DA} \!\Delta t \sum\limits_{s \in \mathscr{S}} p^{-}_{s,t} \!\!+ \hat{\lambda}_t^{DA} \!\Delta t \sum\limits_{s \in \mathscr{S}} p^{+}_{s,t}~; 
     &    
    \forall t \label{MILP_ES: DAMprice} \\
    &{\mu^{SR,\uparrow}} + \xi_t^{SR,\uparrow} \ge ~\check{\lambda}_t^{SR,\uparrow} \sum\limits_{s \in \mathscr{S}} r_{s,t}^{\uparrow}~; 
     &   
    \forall t \label{MILP_ES: upSRMprice} \\
    &{\mu^{SR,\downarrow}} + \xi_t^{SR,\downarrow} \ge ~\check{\lambda}_t^{SR,\downarrow} \sum\limits_{s \in \mathscr{S}} r_{s,t}^{\downarrow}~; 
     &  
    \forall t \label{MILP_ES: downSRMprice} \\
    & C^{\mathrm{S}} \leq 0;
    &    
    \label{MILP_ES: Other_Cons_Det}
\end{align}
\label{Single-level_MILP_ES}
\end{subequations}
\vspace{-2.5em}
\endgroup

\section{Case Studies}
\label{sec:Case_Studies}

\begin{figure} [t!]
    \centering  \includegraphics[width=1\columnwidth]{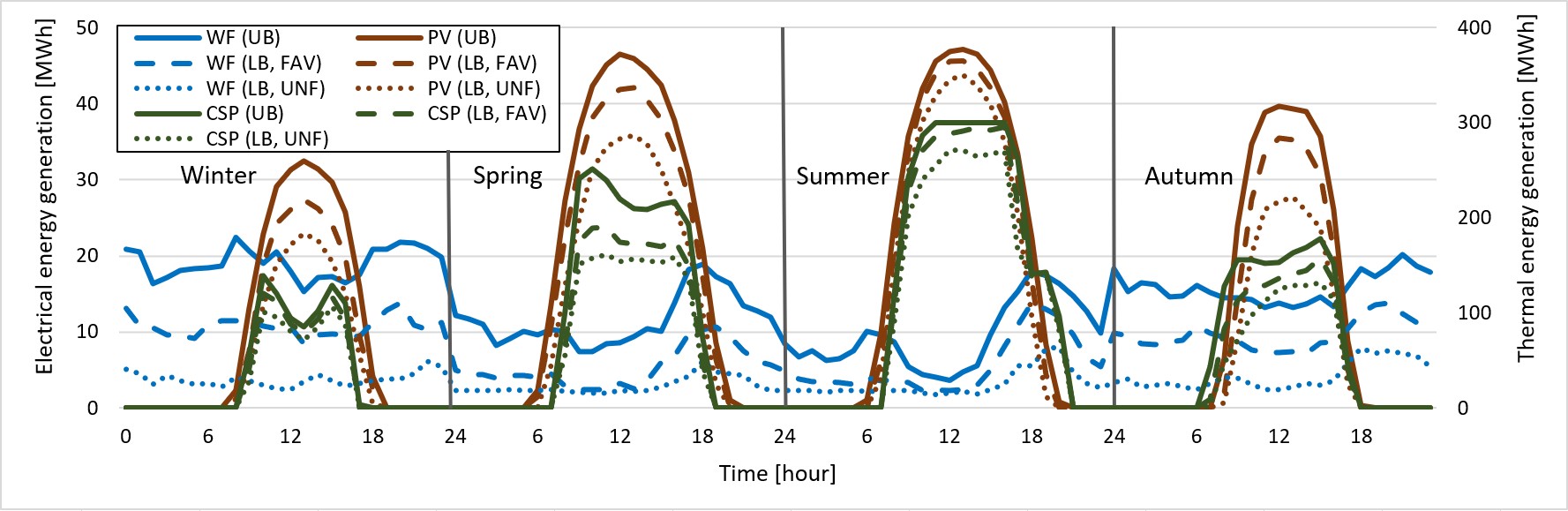}
    \vspace{-2em}
    \caption{The forecast bounds of \ac{wf}, solar PV, and \ac{csp}.}
    \label{fig:Data_Production}
     \vspace{-.7em}
\end{figure}

\begin{figure} [t!]
    \centering  \includegraphics[width=1\columnwidth]{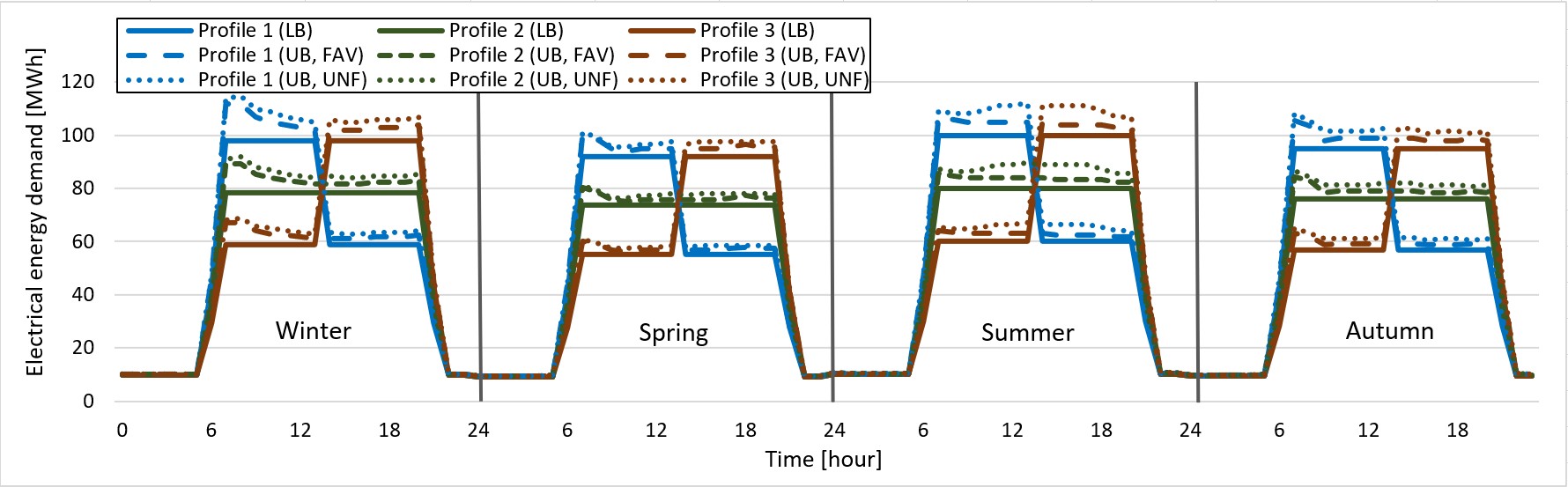}
    \vspace{-1.9em}
    \caption{The forecast bounds of different profiles of \ac{fd}.}
    \label{fig:Data_Demand}
   \vspace{-.7em}
\end{figure}

\begin{figure} [t!]
    \centering  \includegraphics[width=1\columnwidth]{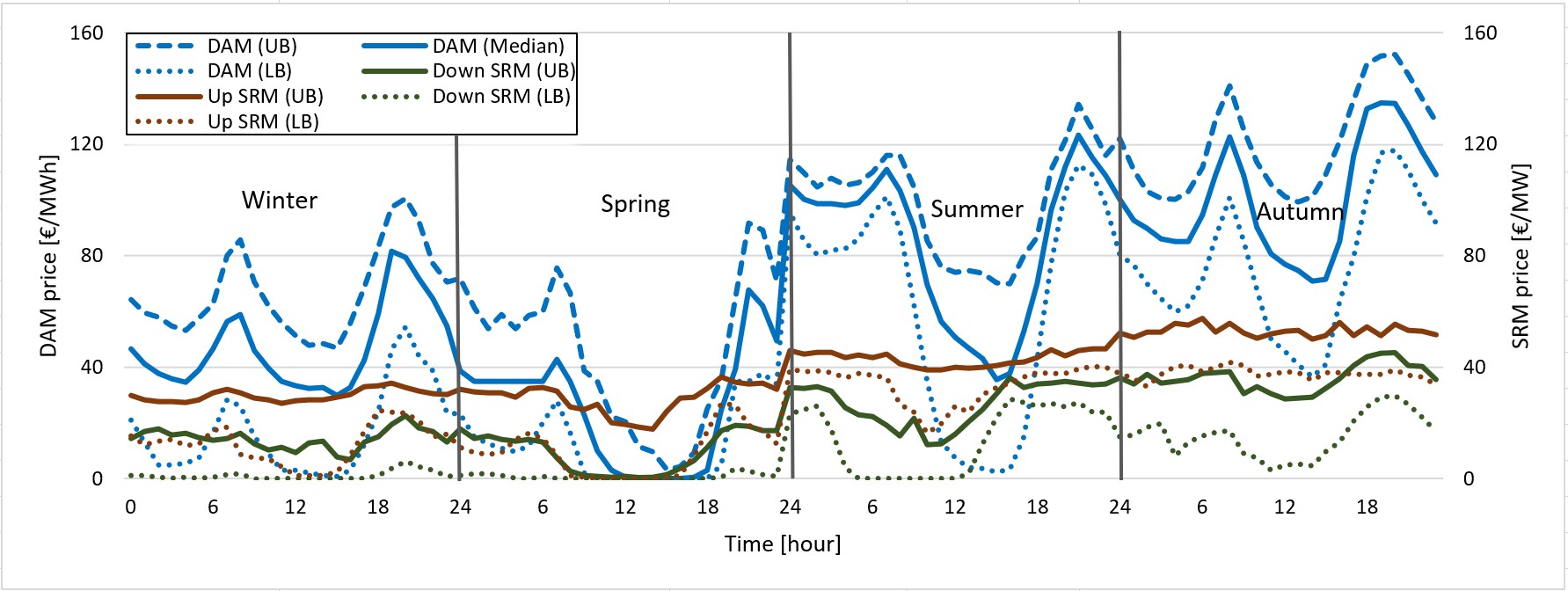}
    \vspace{-2em}
    \caption{The forecast bounds of DAM and SRM price.}
    \label{fig:Data_Price}
    \vspace{-1.1em}
\end{figure}

This section presents the simulation results based on the proposed \ac{ro} framework for evaluating \ac{rvpp} participation in the \ac{dam} and \ac{srm}. Additional simulations assess the sizing and operation of an individual \ac{es}, as well as its integration with \ac{ndrs}, to replicate \ac{rvpp} performance under similar market conditions. The simulations consider an \ac{rvpp} comprising a hydro plant, a biomass unit, a \ac{wf}, a solar \ac{pv} plant, a \ac{csp} equipped with \ac{ts}, and a \ac{fd}. For the electricity market simulations involving the \ac{es}, multiple 1 MWh Li-ion \acp{es} are aggregated to provide the necessary capacity, with their characteristics provided in Table~\ref{table:Data_ES} based on~\cite{ortega2022modeling}. Forecast bounds for the \ac{ndrs} units—including electrical outputs of the \ac{wf} and solar \ac{pv}, and thermal output of the \ac{csp}—are shown in Figure~\ref{fig:Data_Production} across four representative days. These bounds are derived from historical data: solar \ac{pv} and \ac{csp} from CIEMAT Spain~\cite{web:ciemat_spain}, and the \ac{wf} from Iberdrola Spain~\cite{web:iberdrola_spain}. Figure~\ref{fig:Data_Production} includes the \ac{ub} representing the deterministic forecast, and two \ac{lb} scenarios: \textit{favorable} (FAV), indicating moderate uncertainty, and \textit{unfavorable} (UNF), representing greater forecast variability. Both the \ac{wf} and solar \ac{pv} plants are modeled with nominal capacities of 50 MW, with operating costs of 15 €/MWh and 10 €/MWh, respectively. The technical specifications of \ac{csp} are provided in Table~\ref{table:Data_STU}. Operating costs for all units have been levelized based on estimated operational expenses of different generation technologies in~\cite{IEA_LCOE_Calculator}. Information related to the \ac{drs}, including the hydro and biomass plants, is drawn from~\cite{ortega2022modeling} and consolidated in Table~\ref{table:Data_DRES}. The seasonal energy limits of the hydro plant are based on the historical scheduling of units with water reservoir constraints and are set at 1164, 972, 528, and 708 MWh for favorable representative days in winter, spring, summer, and autumn, respectively, according to~\cite{REE2025}. Accordingly, under unfavorable condition, these values are reduced to 804, 624, 420, and 612 MWh, respectively. The forecast boundaries for the \ac{fd} are shown in Figure~\ref{fig:Data_Demand}, using three representative demand profiles from~\cite{ortega2022modeling}. Each base demand profile includes a 10\% flexibility margin, and the \ac{ub} is specified for both favorable and unfavorable scenarios. Additionally, the forecast bounds for electricity prices in the \ac{dam} and \ac{srm} are based on historical data from~\cite{REE2025} and visualized in Figure~\ref{fig:Data_Price}. Table~\ref{table:Data_Budget} presents the uncertainty budgets associated with various uncertain parameters. Since the solar \ac{pv} production, thermal output of the \ac{sf} are zero at night, and demand fluctuations are minimal, these uncertainty budgets are assigned smaller numbers. This allocation strategy maintains a consistent proportion of uncertain hours across the simulation horizon for all parameters. The simulation analysis comprises four case studies to assess the performance of the proposed models for both the \ac{rvpp} and the \ac{es} as follows:
\begin{itemize}
    \item \text{Case 1:} Examine the optimal operation of \ac{rvpp} units, and the \ac{rvpp}'s trading strategy in the \ac{dam} and \ac{srm} under an optimistic strategy over different sample days.

    \item \text{Case 2:} Evaluate the optimal trading strategy of the \ac{rvpp} under different scheduling regimes: \textit{favorable} (moderate energy limits for the hydro plant and moderate forecast variation in solar \ac{pv}, \ac{csp}, \ac{wf} production, and demand consumption), and \textit{unfavorable} (strict energy limits and high forecast variation), across three uncertainty-handling strategies: optimistic, balanced, and pessimistic.

    \item \text{Case 3:} Assess the value of different \ac{rvpp} configurations compared to the individual participation of units in the market. Additionally, evaluate the \ac{es} required to match the \ac{rvpp}'s performance under different uncertainty-handling and \ac{es} integration strategies.

    \item \text{Case 4:} Evaluate the trading strategy of the \ac{es} required to match the performance of the \ac{rvpp} under different uncertainty-handling strategies.

\end{itemize}

\begin{table}[t!]
  \centering
  \caption{Li-ion \ac{es} data.}
   \footnotesize
  \setlength{\tabcolsep}{1pt}
  \renewcommand{\arraystretch}{.7}
  \vspace{-1em}
  \begin{threeparttable}
  \begin{tabular}{lcc}
    \toprule

    \multicolumn{1}{c}{\textbf{Parameter}}  
    & \multicolumn{1}{c}{\textbf{ESS}} 
    \\

    \cmidrule{1-2} 

   \multirow{1}{*}{Charging/discharging power [MW]}    & 0.5/0.5   \\ [0.2em]

    \multirow{1}{*}{\text{Maximum/minimum energy [MWh]}}    & 1/0.1  \\ [0.2em] 

      \multirow{1}{*}{\text{Degradation and operational costs [€/MWh]}}   & 30  \\ [0.2em] 


    \multirow{1}{*}{Charging/discharging efficiency [\%]}   & 95 \\ [0.2em]

\bottomrule
  \end{tabular}
\end{threeparttable}
  \label{table:Data_ES}
      \vspace{-1.5em}
\end{table}

\begin{table}[t!]
  \centering
  \caption{\ac{csp} data.}
  \footnotesize
  \setlength{\tabcolsep}{1pt}
  \renewcommand{\arraystretch}{.8}
  \vspace{-1em}
  \begin{threeparttable}
  \begin{tabular}{lc}
    \toprule

    \multicolumn{1}{c}{\textbf{Parameter}}  
    & \multicolumn{1}{c}{\textbf{Value}}
    \\

    \cmidrule{1-2} 
   \multirow{1}{*}{\text{\ac{sf} maximum thermal power output [MW]}}    & 300 \\ [0.2em]

    \multirow{1}{*}{\text{Turbine maximum thermal power input [MW]}}    & 140 \\ [0.2em] 

    \multirow{1}{*}{\text{Turbine maximum electrical power output [MW]}}    & 55 \\ [0.2em] 

    \multirow{1}{*}{\text{Turbine minimum up/down time [hour]}}    & 3/2 \\ [0.2em]

    \multirow{1}{*} \text{\ac{ts} maximum/minimum thermal energy [MWh]}
   & 1100/110    \\ [0.2em]


    \multirow{1}{*}\text{\ac{ts} charging/discharging thermal power [MW] }   & 140/115   \\ [0.2em] 

    \multirow{1}{*} \text{\ac{ts} charging/discharging efficiency [\%]}   & 95    \\ [0.2em]

        \multirow{1}{*} \text{\ac{csp} operation cost [€/MWh]}
   & 25   \\ [0.2em]

\bottomrule
  \end{tabular}
\end{threeparttable}
  \label{table:Data_STU}
      \vspace{-2em}
\end{table}

\begin{table}[t!]
  \centering
  \caption{\ac{drs} data.}
  \footnotesize
  \setlength{\tabcolsep}{1pt}
  \renewcommand{\arraystretch}{.7}
  \vspace{-1em}
  \begin{threeparttable}
  \begin{tabular}{lccccc}
    \toprule

    \multicolumn{1}{c}{\textbf{Parameter}}  
    && \multicolumn{1}{c}{\textbf{Hydro}}
    && \multicolumn{1}{c}{\textbf{Biomass}} 
    \\

    \cmidrule{1-1} \cmidrule{3-3} \cmidrule{5-5}

   \multirow{1}{*}{\text{Maximum/minimum power output [MW]}}    && 50/10 
   && 10/2  \\ [0.2em]

    \multirow{1}{*}{\text{Startup/shutdown cost [€]}}    && 100/50 &&  300/150  \\ [0.2em] 

   \multirow{1}{*}{\text{Operation cost [€/MWh]}}   && 12.5 &&  60   \\ [0.2em] 

     \multirow{1}{*}{\text{Minimum up/down time [hour]}}    && 1/0 && 3/3    \\ [0.2em]


\bottomrule
  \end{tabular}
\end{threeparttable}
  \label{table:Data_DRES}
      \vspace{-1.3em}
\end{table}

\begin{table}[t!]
  \centering
  \caption{Uncertainty budgets for the \ac{rvpp} operator’s strategies.}
  \footnotesize
  \setlength{\tabcolsep}{.8pt}
  \renewcommand{\arraystretch}{.7}
  \vspace{-1em}
  \begin{threeparttable}
  \begin{tabular}{lcccccc}
    \toprule

    \multicolumn{1}{c}{\textbf{}}   
    && \multicolumn{1}{c}{\textbf{\ac{dam}/\ac{srm}}} 
    & \multicolumn{1}{c}{\textbf{\ac{wf}}}
    & \multicolumn{1}{c}{\textbf{PV}} 
    & \multicolumn{1}{c}{\textbf{\ac{sf} thermal}}
    & \multicolumn{1}{c}{\textbf{\ac{fd}}}
    \\

    \multicolumn{1}{c}{\textbf{Strategy}}   
    && \multicolumn{1}{c}{\textbf{price}} 
    & \multicolumn{1}{c}{\textbf{production}}
    & \multicolumn{1}{c}{\textbf{production}} 
    & \multicolumn{1}{c}{\textbf{production}}
    & \multicolumn{1}{c}{\textbf{consumption}}
    \\

 \cmidrule{1-1} \cmidrule{3-7}

    \multirow{1}{*}{Optimistic}  && 3 & 3  &  2 & 2 & 2  \\ [0.2em]

    \multirow{1}{*}{Balanced}    && 6 & 6   & 4 & 4 &  4 \\ [0.2em]

    \multirow{1}{*}{Pessimistic}    && 9 & 9  & 6 & 6 & 6 \\ [0.2em]

\bottomrule
  \end{tabular}
\end{threeparttable}
  \label{table:Data_Budget}
      \vspace{-1.5em}
\end{table}

The simulations are executed on a Dell XPS with an i7-1165G7 2.8 GHz processor and 16 GB RAM, using the CPLEX solver in GAMS 49. In all simulations, solution time stays under five minutes, highlighting the proposed model’s efficiency in solving multi-market scheduling problems.

\subsection{Case 1}
\label{subsec: Case 1}

\begin{figure} [t!]
    \centering  \includegraphics[width=1\columnwidth]{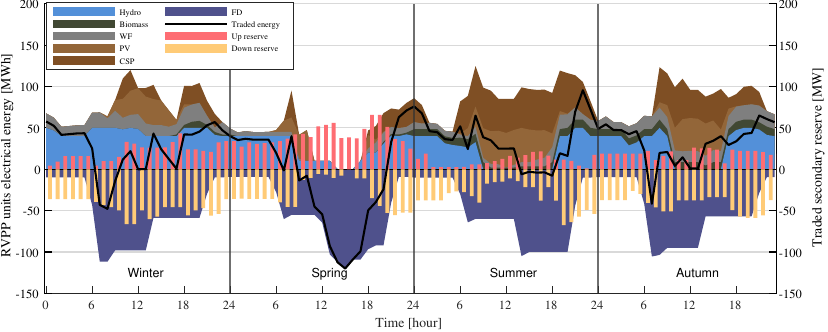}
    \vspace{-2em}
    \caption{Electrical energy generation by RVPP units and traded energy and reserve by the RVPP (Case 1).}
    \label{fig:Units_energy_Case1}
    \vspace{-.5em}
\end{figure}

Figure~\ref{fig:Units_energy_Case1} illustrates the energy scheduling of the \ac{rvpp} units, and the energy and reserve traded by the \ac{rvpp} in the electricity markets under favorable condition and the optimistic strategy. The results indicate that the \ac{rvpp} effectively schedules its units based on the availability of their production across seasons. In winter, the highest share of the \ac{rvpp}'s energy comes from the hydro plant, while solar \ac{pv} and \ac{csp} generation is lower due to limited solar availability. In contrast, during the summer, most of the \ac{rvpp}'s energy is provided by solar \ac{pv} and \ac{csp}, and hydro generation is mostly limited to the morning and night hours. In spring, due to low electricity prices during hours 12–18, a significant portion of the \ac{rvpp}'s production is curtailed, and the \ac{rvpp} primarily supplies its demand by purchasing energy from the market. During these hours, the \ac{rvpp} has limited capacity to provide down reserve due to reduced production, while its capacity to provide up reserve is increased. In autumn, although the available energy from solar \ac{pv} and \ac{csp} is lower compared to summer, this shortfall is compensated by other \ac{rvpp} units such as the \ac{wf} and the hydro plant. Different profiles of \ac{fd} are selected across the seasons (profile 1 in winter and autumn, and profile 3 in spring and summer) to maximize the profitability of the \ac{rvpp}. In winter and autumn, due to higher energy demand during the morning hours, the \ac{rvpp} acts as an energy buyer during hours 8–10 and hour 8, respectively. The selection of profile 1, which features lower demand during the late hours, allows the \ac{rvpp} to exploit higher electricity prices in the afternoon and evening by selling excess energy to the market. In contrast, profile 3, with higher late-hour demand, is optimal in spring and summer due to low afternoon prices (hours 12–18 in Figure~\ref{fig:Data_Price}), enabling the \ac{rvpp} to sell energy in the morning and increase profitability. Furthermore, the biomass plant primarily generates electricity during the evening and/or morning hours—periods when either electricity prices are higher or the \ac{rvpp} lacks sufficient production from other units. This operational strategy is consistent across most seasons, as the biomass plant has a higher production cost compared to other generating units.

\begin{figure} [t!]
    \centering  \includegraphics[width=1\columnwidth]{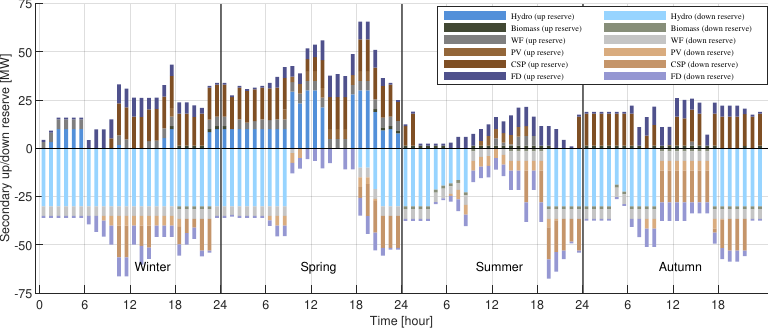}
    \vspace{-1.8em}
    \caption{Up and down reserve provided by RVPP units (Case 1).}
    \label{fig:Units_reserve_Case1}
    \vspace{-1.4em}
\end{figure}

Figure~\ref{fig:Units_reserve_Case1} shows each \ac{rvpp} unit’s contribution to up and down reserve across seasons. The results indicate that the \ac{rvpp} units are effectively scheduled to provide reserves, enhancing the overall profitability of the \ac{rvpp}. Specifically, both the \ac{csp}—owing to the flexibility from its \ac{ts}—and the \ac{fd} contribute significantly to up and down reserve provision in most seasons. The hydro plant is mainly used for energy generation but also offers capacity for up reserve in certain seasons, particularly spring and partially in winter. Moreover, it plays a substantial role in providing down reserve, especially in winter when its production is at its peak. Other units, including the \ac{wf}, solar \ac{pv}, and biomass plant, contribute marginally to reserve provision due to limited flexibility, lower reserve capability, or lower energy production capacity.

\subsection{Case 2}
\label{subsec: Case 2}

\begin{figure} [t!]
    \centering  \includegraphics[width=1\columnwidth]{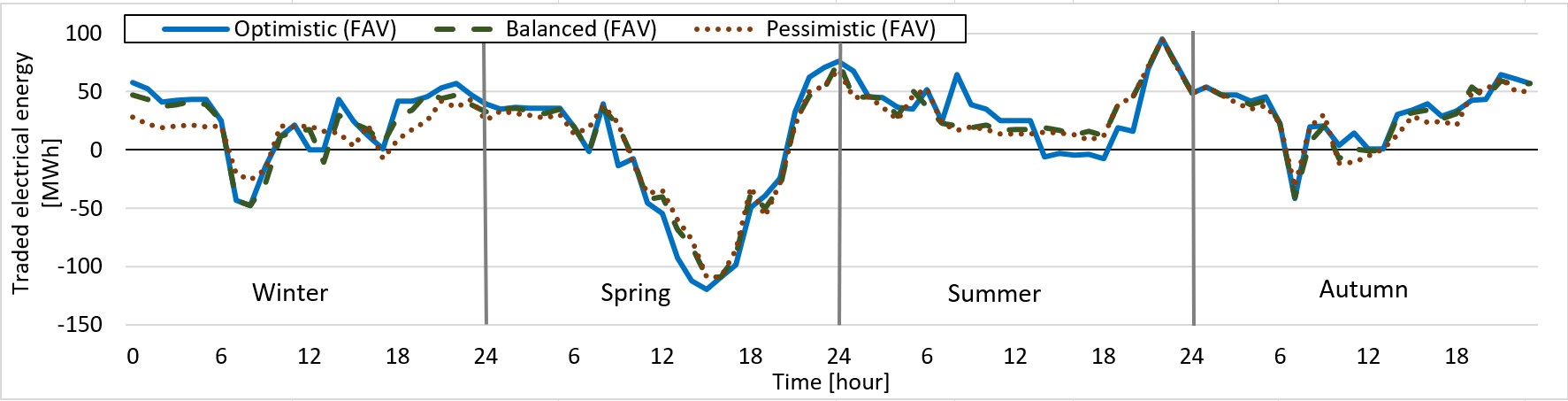}
    \vspace{-.5em}
    \includegraphics[width=1\columnwidth]{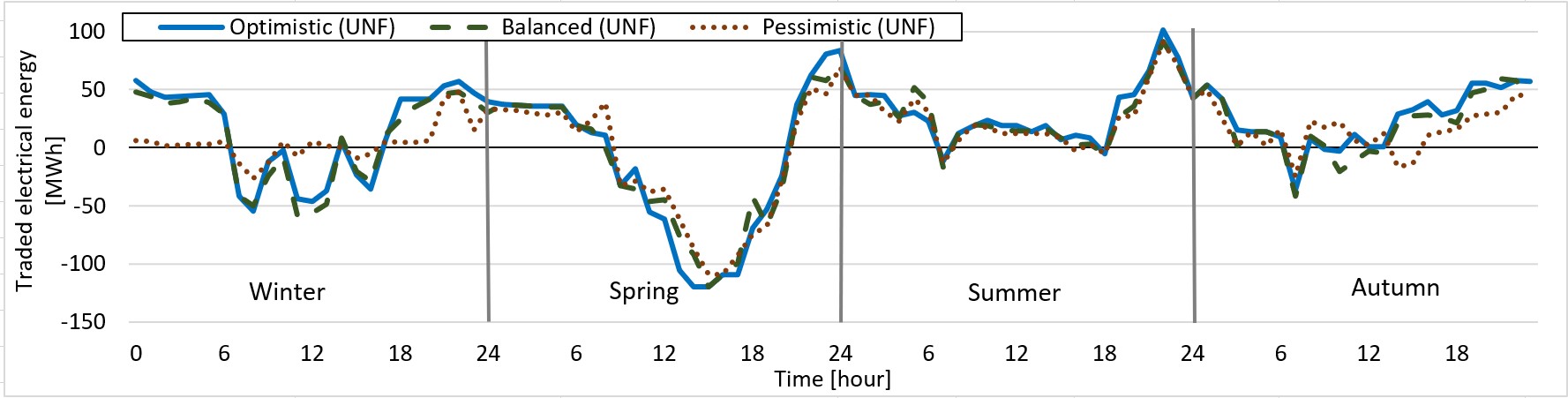}
    \caption{Electrical energy traded by the RVPP (Case 2).}
    \label{fig:RVPP_trade_case2}
    \vspace{-1.2em}
\end{figure}

Figure~\ref{fig:RVPP_trade_case2} presents the electrical energy traded by the \ac{rvpp} under different scheduling regimes (favorable and unfavorable) and uncertainty-handling strategies (optimistic, balanced, and pessimistic). The results show that, under favorable condition, the \ac{rvpp} tends to reduce energy sales in most hours and seasons as more conservative strategies (i.e., balanced and pessimistic) are adopted to manage uncertainty. However, this trend is not uniform across all hours, due to the \ac{rvpp}'s flexibility in selecting different \ac{fd} profiles. For instance, in summer, the balanced and pessimistic strategies result in the selection of profile 2 of \ac{fd}, whereas the optimistic strategy selects profile 3, which has higher demand during late hours. Consequently, the \ac{rvpp} trades more energy between hours 15–21 under the balanced and pessimistic strategies than in the optimistic case. Under unfavorable condition, the \ac{rvpp}'s energy trading is constrained in more hours, particularly when employing the pessimistic strategy. For example, during winter under the unfavorable-pessimistic scenario, the energy sold by the \ac{rvpp} is nearly zero throughout most hours, except for a limited period between hours 22–24. In autumn, high variability in generation—especially from solar-dependent units—leads to significant changes in the \ac{rvpp}'s trading behavior. By contrast, in summer, solar generation is less affected under unfavorable condition during daylight hours, allowing for more stable trading patterns. However, in hour 8, the \ac{rvpp} acts as an energy buyer across all uncertainty-handling strategies under unfavorable condition, whereas it operates as an energy seller in the favorable case.

\begin{figure} [t!]
    \centering  \includegraphics[width=1\columnwidth]{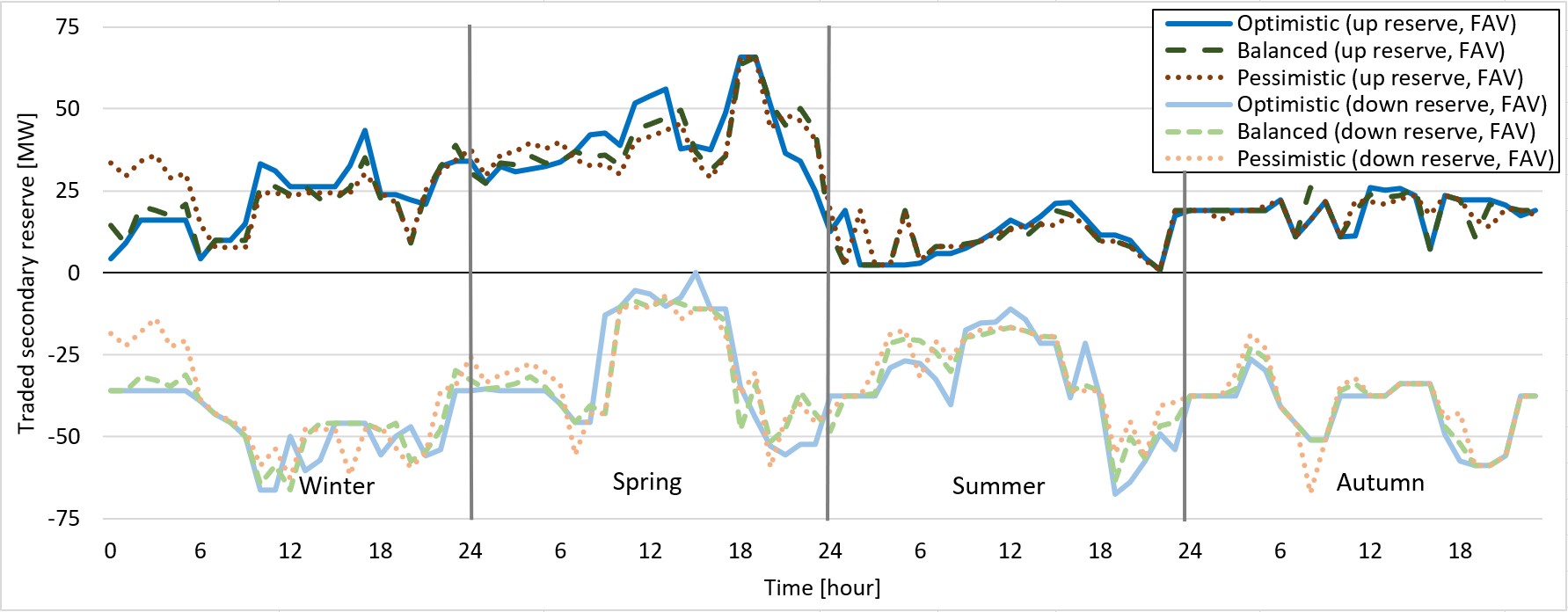}
    \vspace{-.5em}
    \includegraphics[width=1\columnwidth]{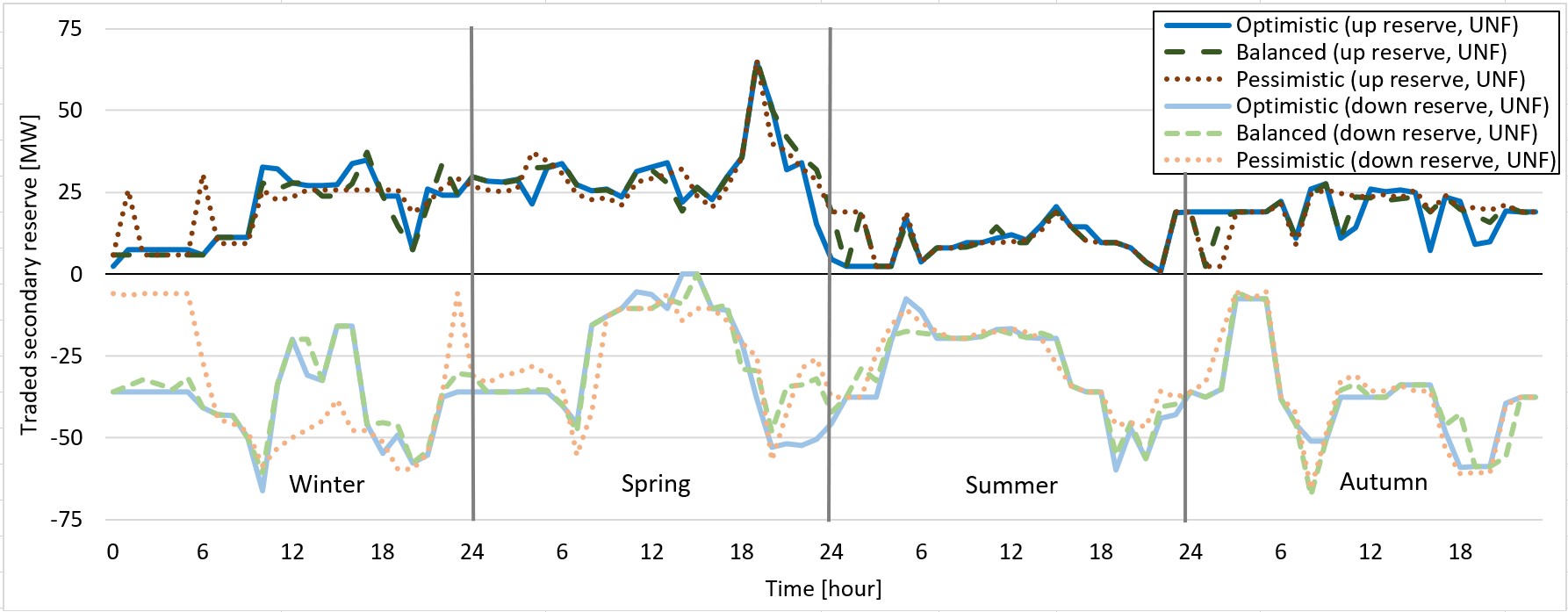}
    \caption{Up and down reserves traded by the RVPP (Case 2).}
    \label{fig:RVPP_reserve_case2}
    \vspace{-.8em}
\end{figure}

Figure~\ref{fig:RVPP_reserve_case2} illustrates the up and down reserves traded by the \ac{rvpp} in the \ac{srm} under different scheduling regimes and uncertainty-handling strategies. The reserve capacity provided by the \ac{rvpp} is influenced by both its traded energy and the level of uncertainty. Under favorable condition, particularly during winter in hours 1–6, up reserve provision increases in the balanced and pessimistic strategies compared to the optimistic one. This is primarily because the \ac{rvpp} schedules less energy for market sale during these hours (see Figure~\ref{fig:RVPP_trade_case2}), thereby retaining more capacity to offer up reserves. However, this pattern is not consistent across all hours. For example, in hour 20 of winter, both traded energy and up reserve are reduced in the balanced and pessimistic strategies compared to the optimistic strategy. Similar trends are observed for down reserve provision, where the reserve levels vary depending on the chosen uncertainty-handling strategy. In the unfavorable scenario, the \ac{rvpp}'s ability to provide down reserves is more significantly impacted than up reserves—both in comparison to the favorable condition and across strategies. This is mainly due to the reduced generation availability from \ac{rvpp} units. For instance, during winter in hours 1–5, the down reserve provided under the pessimistic strategy is substantially lower than under the optimistic and balanced strategies, primarily due to minimal or zero output from the hydro plant. Conversely, during hours 12–17, the hydro plant is dispatched under the pessimistic strategy (but not under the optimistic and balanced ones), enabling greater down reserve provision in the pessimistic case for those hours.

\subsection{Case 3}
\label{subsec: Case 3}

Table~\ref{table:RVPP_profit_Case3} presents a comparison of the profit generated by each \ac{rvpp} unit when participating individually in the \ac{dam} and \ac{srm}, along with the total profit from individual participation of all units and the profit achieved by the \ac{rvpp} under different scheduling regimes and uncertainty-handling strategies. The results indicate that as more conservative strategies are adopted (i.e., balanced and pessimistic), the individual profits of units generally decrease—or their costs increase—under both favorable and unfavorable conditions. For instance, under favorable condition and individual participation, the profit reduction (or cost increase) in the pessimistic strategy compared to the optimistic one is observed to be 15.5\%, 15.5\%, 24.3\%, 44.0\%, 20.8\%, and 16.9\% for the hydro, biomass, \ac{wf}, solar \ac{pv}, \ac{csp}, and \ac{fd} units, respectively. These reductions are more pronounced under unfavorable condition, reaching 17.3\%, 15.5\%, 38.9\%, 48.9\%, 21.3\%, and 18.3\%. When the units participate collectively as an \ac{rvpp}, the resulting profits exceed the sum of individual profits by 65.34/64.31 k€, 112.41/110.58 k€, and 144.49/144.20 k€ under optimistic, balanced, and pessimistic strategies in favorable/unfavorable condition, respectively. These values serve as the \ac{lb} for the \ac{es} sizing problem, which aims to achieve equivalent economic performance to that of the \ac{rvpp} problem. For example, the required energy storage capacities to match the \ac{rvpp} profit under favorable condition are 63, 124, and 177~MWh for the optimistic, balanced, and pessimistic strategies, respectively. Since the \ac{rvpp} becomes more beneficial as uncertainty increases, the required capacity of \ac{es} must also increase to achieve a similar performance as the \ac{rvpp}.

\begin{table}[t!]
  \centering
  \caption{Individual versus aggregated Profit of \ac{rvpp} Units under FAV (green) and UNF (orange) scheduling regimes (Case 3).}
  \scriptsize
  \setlength{\tabcolsep}{2pt}
  \renewcommand{\arraystretch}{.8}
  \vspace{-1.2em}
  \begin{threeparttable}
  \begin{tabular}{clccc}
    \toprule

     \multicolumn{1}{c}{\textbf{}}   
    && \multicolumn{3}{c}{\textbf{Profit (cost) [k€]}} 
    \\

 \cmidrule{1-1} \cmidrule{3-5}

    \multicolumn{1}{c}{\textbf{Unit}}   
    && \multicolumn{1}{c}{\textbf{Optimistic}} 
    & \multicolumn{1}{c}{\textbf{Balanced}}
    & \multicolumn{1}{c}{\textbf{Pessimistic}} 
    \\

 \cmidrule{1-1} \cmidrule{3-5}

\multirow{1}{*}{Hydro} && \textcolor{dgreen}{208.87}/\textcolor{orange}{172.81}  & \textcolor{dgreen}{191.69}/\textcolor{orange}{156.69}  & \textcolor{dgreen}{176.49}/\textcolor{orange}{142.85}  \\ [0.2em]

\multirow{1}{*}{Biomass}  && \textcolor{dgreen}{17.35}/\textcolor{orange}{17.35}  & \textcolor{dgreen}{15.89}/\textcolor{orange}{15.89}  & \textcolor{dgreen}{14.65}/\textcolor{orange}{14.65}  \\ [0.2em] 

\multirow{1}{*}{\ac{wf}}  && \textcolor{dgreen}{68.59}/\textcolor{orange}{65.91}  & \textcolor{dgreen}{58.73}/\textcolor{orange}{52.16}  & \textcolor{dgreen}{51.89}/\textcolor{orange}{40.28}  \\ [0.2em] 

\multirow{1}{*}{\ac{pv}} && \textcolor{dgreen}{39.51}/\textcolor{orange}{38.07}  & \textcolor{dgreen}{27.83}/\textcolor{orange}{25.91}  & \textcolor{dgreen}{22.11}/\textcolor{orange}{19.43}  \\ [0.2em] 

\multirow{1}{*}{\ac{csp}} && \textcolor{dgreen}{117.13}/\textcolor{orange}{114.34}  & \textcolor{dgreen}{101.53}/\textcolor{orange}{99.27}  & \textcolor{dgreen}{92.80}/\textcolor{orange}{90.02}  \\ [0.2em] 

\multirow{1}{*}{\ac{fd}}  && \textcolor{dgreen}{-301.38}/\textcolor{orange}{-301.87}  & \textcolor{dgreen}{-329.40}/\textcolor{orange}{-332.16}  & \textcolor{dgreen}{-352.41}/\textcolor{orange}{-357.08}  \\ [0.2em] 

 \cmidrule{1-5}
 
\multirow{1}{*}{\textbf{Total}}    && \textcolor{dgreen}{150.07}/\textcolor{orange}{106.61}  & \textcolor{dgreen}{66.27}/\textcolor{orange}{17.76}  & \textcolor{dgreen}{5.53}/\textcolor{orange}{-49.85}  \\ [0.2em] 

\multirow{1}{*}{\textbf{\ac{rvpp}}}   && \textcolor{dgreen}{215.41}/\textcolor{orange}{170.92}  & \textcolor{dgreen}{178.68}/\textcolor{orange}{128.34}  & \textcolor{dgreen}{150.02}/\textcolor{orange}{94.35}  \\ [0.2em] 

\bottomrule
  \end{tabular}
\end{threeparttable}
  \label{table:RVPP_profit_Case3}
     \vspace{-2em}
\end{table}

Table~\ref{table:RVPP_wo_Units_profit_Case3} presents the additional profit of various \ac{rvpp} configurations compared to individual participation, under different scheduling regimes and uncertainty-handling strategies. It also presents the required \ac{es} size to match the \ac{rvpp}'s performance for each configuration. In each configuration, one unit technology is excluded from the \ac{rvpp} to better assess the contribution of each technology. The results reveal that the value of each unit within the \ac{rvpp} depends on several factors, such as its capacity, available production, and dispatchability. In the configurations without the biomass plant and \ac{wf}, the additional profit under optimistic and favorable conditions is reduced by only 1\% and 4.9\%, respectively, compared to the full \ac{rvpp}. This is because these units either have a low capacity (biomass plant) or limited available production (\ac{wf}) compared to other \ac{rvpp} units. Conversely, excluding the \ac{fd} from the \ac{rvpp} results in a significant 78.5\% reduction in additional profit. The reason is that a considerable share of the energy produced by the \ac{rvpp} is consumed by its internal demand. Including the \ac{fd} in the \ac{rvpp} significantly increases profitability due to its flexibility and its role in reducing energy spillage from \ac{ndrs}. The \ac{csp} and hydro plants provide considerable value within the \ac{rvpp} and are crucial for its profitability. When excluded, the additional \ac{rvpp} profit decreases by 20.2\% and 23.4\%, respectively. This is because the \ac{csp}, supported by its \ac{ts}, mitigates input thermal energy uncertainty and offers dispatchable energy, while the hydro plant is inherently dispatchable.


\begin{table*}[t!]
  \centering
  \caption{Additional profit of different \ac{rvpp} configurations and corresponding \ac{es} size to match \ac{rvpp} performance (Case 3).}
  \scriptsize
  \setlength{\tabcolsep}{2pt}
  \renewcommand{\arraystretch}{.8}
  \vspace{-1.2em}
  \begin{threeparttable}
  \begin{tabular}{clccclccc}
    \toprule

     \multicolumn{1}{c}{\textbf{}}   
    && \multicolumn{3}{c}{\textbf{RVPP additional profit [k€]}} 
    && \multicolumn{3}{c}{\textbf{ESS maximum energy [MWh]}} 
    \\

 \cmidrule{1-1} \cmidrule{3-5} \cmidrule{7-9}

    \multicolumn{1}{c}{\textbf{Configuration}}   
    && \multicolumn{1}{c}{\textbf{Optimistic}} 
    & \multicolumn{1}{c}{\textbf{Balanced}}
    & \multicolumn{1}{c}{\textbf{Pessimistic}} 
     && \multicolumn{1}{c}{\textbf{Optimistic}} 
    & \multicolumn{1}{c}{\textbf{Balanced}}
    & \multicolumn{1}{c}{\textbf{Pessimistic}} 
    \\

 \cmidrule{1-1} \cmidrule{3-5} \cmidrule{7-9}


\multirow{1}{*}{\ac{rvpp}}  && \textcolor{dgreen}{65.34}/\textcolor{orange}{64.31}  & \textcolor{dgreen}{112.41}/\textcolor{orange}{110.58}  & \textcolor{dgreen}{144.49}/\textcolor{orange}{144.20}  && \textcolor{dgreen}{63}/\textcolor{orange}{62}  & \textcolor{dgreen}{124}/\textcolor{orange}{121}  & \textcolor{dgreen}{177}/\textcolor{orange}{176}  \\ [0.2em] 

\multirow{1}{*}{\ac{rvpp} w/o Hydro} && \textcolor{dgreen}{50.03}/\textcolor{orange}{49.62}  & \textcolor{dgreen}{89.37}/\textcolor{orange}{86.74}  & \textcolor{dgreen}{115.56}/\textcolor{orange}{110.33} && \textcolor{dgreen}{48}/\textcolor{orange}{48}  & \textcolor{dgreen}{98}/\textcolor{orange}{95}  & \textcolor{dgreen}{141}/\textcolor{orange}{136}  \\ [0.2em] 

\multirow{1}{*}{\ac{rvpp} w/o Biomass} && \textcolor{dgreen}{64.65}/\textcolor{orange}{63.55}  & \textcolor{dgreen}{111.44}/\textcolor{orange}{109.37}  & \textcolor{dgreen}{143.37}/\textcolor{orange}{142.31}  && \textcolor{dgreen}{62}/\textcolor{orange}{61}  & \textcolor{dgreen}{122}/\textcolor{orange}{120}  & \textcolor{dgreen}{175}/\textcolor{orange}{174}  \\ [0.2em] 

\multirow{1}{*}{\ac{rvpp} w/o \ac{wf}} && \textcolor{dgreen}{62.13}/\textcolor{orange}{61.52}  & \textcolor{dgreen}{107.25}/\textcolor{orange}{105.53}  & \textcolor{dgreen}{137.84}/\textcolor{orange}{135.67}  && \textcolor{dgreen}{60}/\textcolor{orange}{59}  & \textcolor{dgreen}{117}/\textcolor{orange}{115}  & \textcolor{dgreen}{170}/\textcolor{orange}{166}  \\ [0.2em] 

\multirow{1}{*}{\ac{rvpp} w/o \ac{pv}} && \textcolor{dgreen}{52.41}/\textcolor{orange}{50.42}  & \textcolor{dgreen}{89.76}/\textcolor{orange}{87.56}  & \textcolor{dgreen}{116.99}/\textcolor{orange}{115.34}  && \textcolor{dgreen}{51}/\textcolor{orange}{49}  & \textcolor{dgreen}{98}/\textcolor{orange}{96}  & \textcolor{dgreen}{143}/\textcolor{orange}{141}  \\ [0.2em] 

\multirow{1}{*}{\ac{rvpp} w/o \ac{csp}} && \textcolor{dgreen}{52.14}/\textcolor{orange}{50.80}  & \textcolor{dgreen}{88.84}/\textcolor{orange}{86.21}  & \textcolor{dgreen}{115.99}/\textcolor{orange}{114.05}  && \textcolor{dgreen}{50}/\textcolor{orange}{49}  & \textcolor{dgreen}{97}/\textcolor{orange}{95}  & \textcolor{dgreen}{142}/\textcolor{orange}{140}  \\ [0.2em] 

\multirow{1}{*}{\ac{rvpp} w/o \ac{fd}} && \textcolor{dgreen}{14.05}/\textcolor{orange}{14.63}  & \textcolor{dgreen}{21.50}/\textcolor{orange}{23.49}  & \textcolor{dgreen}{22.19}/\textcolor{orange}{25.98}  && \textcolor{dgreen}{14}/\textcolor{orange}{14}  & \textcolor{dgreen}{24}/\textcolor{orange}{26}  & \textcolor{dgreen}{28}/\textcolor{orange}{32}  \\ [0.2em]

\bottomrule
  \end{tabular}
\end{threeparttable}
  \label{table:RVPP_wo_Units_profit_Case3}
     \vspace{-2.5em}
\end{table*}

Table~\ref{table:RVPP_wo_FD_profit_Case3} shows the additional profit of the \ac{rvpp} compared to individual participation, assuming different percentages of \ac{fd} capacity relative to the values in Figure~\ref{fig:Data_Demand}. The results indicate that increasing the capacity of the \ac{fd} leads to higher additional profit for the \ac{rvpp}. However, the percentage increase in profit is more substantial at lower \ac{fd} capacities. For instance, when the \ac{fd} capacity is 50\%, the additional profit—under optimistic and favorable conditions—is 203.2\% higher compared to the \ac{rvpp} configuration without \ac{fd}. Increasing the capacity from 50\% to 100\% results in a 53.4\% profit increase, while a further increase from 100\% to 150\% yields an additional 20.8\% profit increase. The primary reason for this trend is that the initial increase in \ac{fd} capacity significantly reduces energy curtailment from \ac{ndrs}. At higher capacities, the additional profit mainly comes from the \ac{fd}'s ability to provide more efficient trading strategies through coordination with the \ac{rvpp} units.

\begin{table}[t!]
  \centering
  \caption{RVPP additional profit compared to individual profits for different capacity of \ac{fd} (Case 3).}
  \scriptsize
  \setlength{\tabcolsep}{2pt}
  \renewcommand{\arraystretch}{.8}
  \vspace{-1.2em}
  \begin{threeparttable}
  \begin{tabular}{clccc}
    \toprule

     \multicolumn{1}{c}{\textbf{}}   
    && \multicolumn{3}{c}{\textbf{Additional profit [k€]}} 
    \\

 \cmidrule{1-1} \cmidrule{3-5}

    \multicolumn{1}{c}{\textbf{FD capacity}}   
    && \multicolumn{1}{c}{\textbf{Optimistic}} 
    & \multicolumn{1}{c}{\textbf{Balanced}}
    & \multicolumn{1}{c}{\textbf{Pessimistic}} 
    \\

 \cmidrule{1-1} \cmidrule{3-5}


\multirow{1}{*}{0\%} && \textcolor{dgreen}{14.05}/\textcolor{orange}{14.63}  & \textcolor{dgreen}{21.50}/\textcolor{orange}{23.49}  & \textcolor{dgreen}{22.19}/\textcolor{orange}{25.98}  \\ [0.2em]  

\multirow{1}{*}{50\%} && \textcolor{dgreen}{42.60}/\textcolor{orange}{41.58}  & \textcolor{dgreen}{72.39}/\textcolor{orange}{72.22}  & \textcolor{dgreen}{91.52}/\textcolor{orange}{91.61}  \\ [0.2em] 

\multirow{1}{*}{100\%} && \textcolor{dgreen}{65.34}/\textcolor{orange}{64.31}  & \textcolor{dgreen}{112.41}/\textcolor{orange}{110.58}  & \textcolor{dgreen}{144.49}/\textcolor{orange}{144.20}  \\ [0.2em] 

\multirow{1}{*}{150\%} && \textcolor{dgreen}{78.92}/\textcolor{orange}{76.40}  & \textcolor{dgreen}{135.88}/\textcolor{orange}{131.95}  & \textcolor{dgreen}{176.23}/\textcolor{orange}{168.39}  \\ [0.2em]

\bottomrule
  \end{tabular}
\end{threeparttable}
  \label{table:RVPP_wo_FD_profit_Case3}
     \vspace{-3em}
\end{table}

Table~\ref{table:ES+WF_PV_profit_Case3} shows the required size of the \ac{es} when it is coordinated with \ac{ndrs}, such as \ac{wf} and solar \ac{pv}, to match the performance of the \ac{rvpp}. The results indicate that, in order to achieve the same percentage increase in profit as the \ac{rvpp}, larger \ac{es} capacities are generally needed, especially under more conservative strategies and unfavorable condition. For example, in the balanced and pessimistic cases compared to the optimistic case, achieving the same percentage profit increase as the \ac{rvpp} requires a 100\% and 200\% increase in the size of the \ac{es} integrated with the \ac{wf}, and a 66.6\% and 141.7\% increase in the size of the \ac{es} integrated with the \ac{pv}, respectively. The findings also reveal that solar \ac{pv} is more economically viable than \ac{wf}, primarily due to its lower energy variability, as a smaller increase in \ac{es} size is sufficient when coordinated with solar \ac{pv} compared to \ac{wf}.

\begin{table}[t!]
  \centering
  \caption{Size of \ac{es} to match the \ac{rvpp}'s performance (Case 3).}
  \scriptsize
  \setlength{\tabcolsep}{2pt}
  \renewcommand{\arraystretch}{.8}
  \vspace{-1.2em}
  \begin{threeparttable}
  \begin{tabular}{clccc}
    \toprule

     \multicolumn{1}{c}{\textbf{}}   
    && \multicolumn{3}{c}{\textbf{Maximum energy [MWh]}} 
    \\

 \cmidrule{1-1} \cmidrule{3-5}

    \multicolumn{1}{c}{\textbf{Unit}}   
    && \multicolumn{1}{c}{\textbf{Optimistic}} 
    & \multicolumn{1}{c}{\textbf{Balanced}}
    & \multicolumn{1}{c}{\textbf{Pessimistic}} 
    \\

 \cmidrule{1-1} \cmidrule{3-5}


\multirow{1}{*}{\ac{es}+\ac{wf}}  && \textcolor{dgreen}{19}/\textcolor{orange}{23}  & \textcolor{dgreen}{37}/\textcolor{orange}{46}  & \textcolor{dgreen}{57}/\textcolor{orange}{69}  \\ [0.2em] 

\multirow{1}{*}{\ac{es}+\ac{pv}} && \textcolor{dgreen}{10}/\textcolor{orange}{12}  & \textcolor{dgreen}{15}/\textcolor{orange}{20}  & \textcolor{dgreen}{20}/\textcolor{orange}{29}  \\ [0.2em] 

\bottomrule
  \end{tabular}
\end{threeparttable}
  \label{table:ES+WF_PV_profit_Case3}
     \vspace{-2.5em}
\end{table}

\subsection{Case 4}
\label{subsec: Case 4}

\begin{figure} [t!]
    \centering  \includegraphics[width=1\columnwidth]{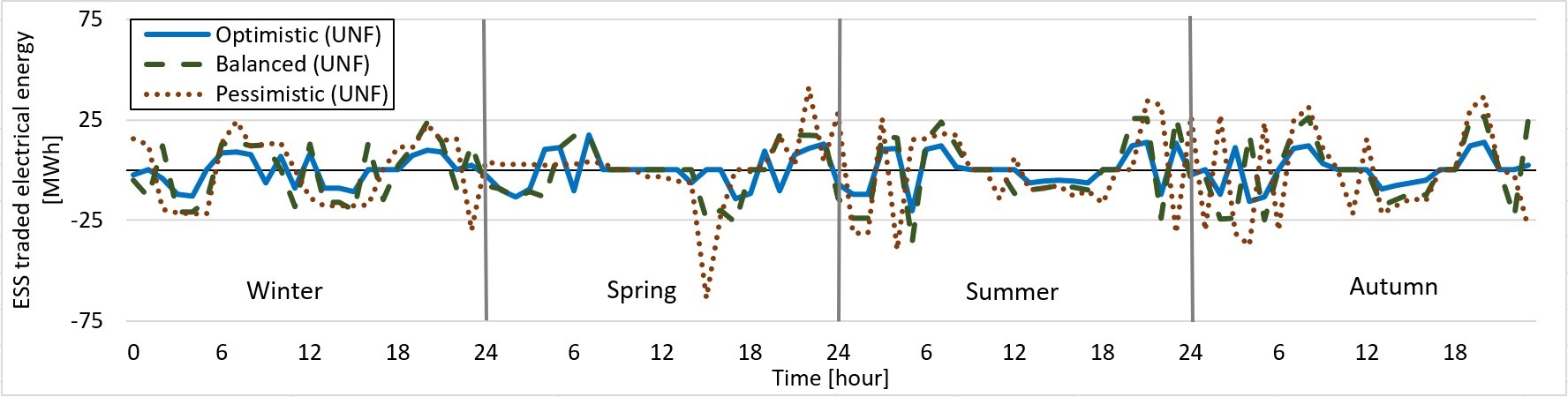}
    \vspace{-1em}
    \includegraphics[width=1\columnwidth]{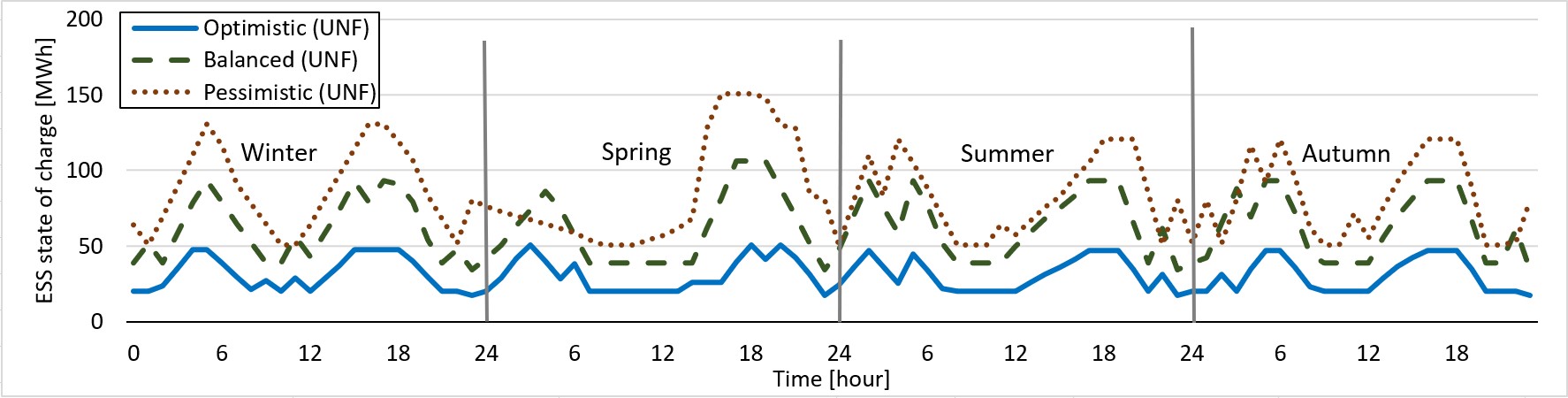}
    \caption{Electrical energy traded by the \ac{es} and its \ac{soc} (Case 4).}
    \label{fig:ES_Energy_case4}
    \vspace{-.5em}
\end{figure}

Figure~\ref{fig:ES_Energy_case4} illustrates the energy traded by the \ac{es} in the \ac{dam} to achieve the same economic performance as the aggregated \ac{rvpp} under different uncertainty-handling strategies in the unfavorable condition. The figure also depicts the corresponding \ac{soc} profile of the \ac{es}. In all strategies, the \ac{es} maximizes market profitability by charging during low-price periods and discharging during high-price periods in the \ac{dam}. Notably, in the pessimistic case, larger fluctuations in both traded energy and the \ac{soc} of the \ac{es} are observed, reflecting the need to manage higher uncertainty more effectively.

\begin{figure} [t!]
    \centering      \includegraphics[width=1\columnwidth]{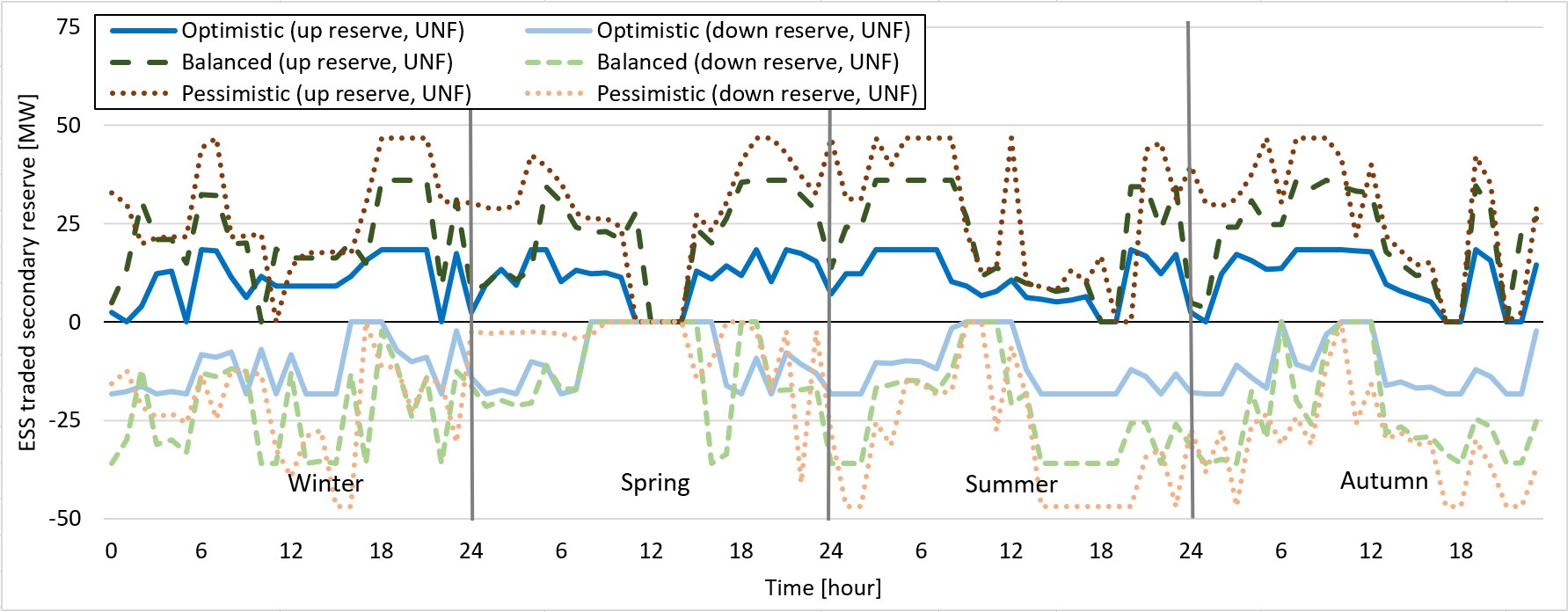}
    \vspace{-2em}
    \caption{Up and down reserves traded by the \ac{es} (Case 4).}
    \label{fig:ES_Reserve_case4}
    \vspace{-1.2em}
\end{figure}

Figure~\ref{fig:ES_Reserve_case4} presents the up and down reserves traded by the \ac{es} in the \ac{srm}. The results indicate that, particularly under the pessimistic strategy—where larger storage capacity is utilized—the provision of both up and down reserves is significantly increased across most hours. Additionally, greater variability in the traded reserve is observed in the pessimistic and balanced strategies compared to the optimistic strategy, reflecting the need to hedge against higher levels of uncertainty. Moreover, the \ac{es} tends to allocate more capacity to reserve provision during hours with higher \ac{srm} prices (compare Figure~\ref{fig:ES_Reserve_case4} with Figure~\ref{fig:Data_Price}), adapting its scheduling across seasons and uncertainty-handling strategies. For instance, in winter under the balanced and pessimistic strategies, a larger share of up reserve is provided during hours 7, 8, and 19–22, corresponding to periods of higher reserve prices.

\section{Conclusion}
\label{sec:Conclusion}

This study compares \acp{rvpp} and grid‑scale \acp{es} for energy and reserve market participation using two‑stage \ac{ro} optimization. Seasonal scheduling regimes and multiple uncertainty sources—prices, generation, and demand—are incorporated to ensure a fair assessment. Four case studies, spanning favorable and unfavorable conditions and three uncertainty‑handling stances (optimistic, balanced, and pessimistic), demonstrate the models’ adaptability. Key insights include:

\begin{itemize}

    \item {
    \textit{Seasonal flexibility}: In winter, the hydro unit supplies most \ac{rvpp} energy, while summer operation relies mainly on solar \ac{pv} and \ac{csp}. The hydro plant dominates down reserve provision, while \ac{csp} and \ac{fd} supply most of the up reserve.}

\item {
    \textit{Uncertainty handling and scheduling adaptation}: More conservative strategies reduce the \ac{rvpp}’s hourly energy sales but portfolio flexibility offsets these reductions, raising total profit. Under unfavorable resource condition, winter and autumn output energy declines sharply, whereas summer remains comparatively stable.}

\item {
    \textit{Economic comparison}: The \ac{rvpp} outperforms individual unit participation, with its profit advantage increasing under higher uncertainty. Specifically, it achieves 72\% and 121.1\% higher profit under balanced and pessimistic strategies, respectively, compared to the optimistic case. While an \ac{es} benefits from energy arbitrage and reserve provision, achieving comparable profitability requires a capacity increase of 96.8\% and 180.9\% under balanced and pessimistic strategies, respectively. Moreover, integrating \ac{es} with solar \ac{pv} is more economically viable than \ac{wf}, mainly due to its lower energy variability.
}

\item {
 \textit{Units' contribution}: 
The contribution of \ac{rvpp} units to overall profitability depends on their energy availability and dispatchability. The \ac{fd}, by reducing energy spillage from \ac{ndrs}, enhances profitability. Furthermore, the \ac{fd}, along with the \ac{csp} and hydro units, significantly improves \ac{rvpp} performance by offering more effective energy and reserve trading options.
}

\end{itemize}




\bibliographystyle{IEEEtran}
\bibliography{refs.bib}


\clearpage

\end{document}